\documentclass[modern, preprint]{aastex62}
\usepackage{graphicx}
\DeclareGraphicsExtensions{.png,.jpg,.eps,.pdf}
\usepackage{natbib}
\usepackage{amsmath}
\usepackage{amsfonts}
\usepackage{amssymb}
\usepackage{lineno}
\usepackage{color}
\submitjournal{ApJ}

\newcommand{\ISOIS}{IS$\odot$IS}

\begin{document}

\title{Parker Solar Probe Observations of Energetic Particles in the Flank of a Coronal Mass Ejection Close to the Sun} 

\author[0000-0002-3737-9283]{N. A. Schwadron}
\affiliation{University of New Hampshire, Durham, NH, 03824, USA}
\affiliation{Department of Astrophysical Sciences, Princeton University, Princeton, NJ, 08544, USA}

\author[0000-0002-1989-3596]{Stuart D. Bale}
\affil{Physics Department, University of California, Berkeley, CA 94720-7300, USA}
\affil{Space Sciences Laboratory, University of California, Berkeley, CA 94720-7450, USA}

\author{J. Bonnell}
\affiliation{University of California at Berkeley, Berkeley, CA 94720, USA}

\author{A. Case}
\affiliation{Harvard-Smithsonian Center for Astrophysics, Cambridge, MA 02138, USA}

\author{M. Shen}
\affiliation{Department of Astrophysical Sciences, Princeton University, Princeton, NJ, 08544, USA}

\author{E. R. Christian}
\affiliation{Goddard Space Flight Center, Greenbelt, MD 20771, USA}

\author{C. M. S. Cohen}
\affiliation{California Institute of Technology, Pasadena, CA 91125, USA}

\author{A. J.  Davis}
\affiliation{California Institute of Technology, Pasadena, CA 91125, USA}

\author{M. I. Desai}
\affiliation{University of Texas at San Antonio, San Antonio, TX 78249, USA }

\author{K. Goetz}
\affiliation{University of Minnesota, Minneapolis, MN 55455, USA}

\author{J. Giacalone}
\affiliation{University of Arizona, Tucson, AZ 85721, USA }

\author{M. E. Hill}
\affiliation{Applied Physics Laboratory, Laurel, MD 20723, USA}


\author{J. C. Kasper}
\affiliation{University of Michigan, Ann Arbor, MI 48109, USA}

\author{K. Korreck}
\affiliation{NASA HQ, Science Mission Directorate, Heliophysics Division, Mary W. Jackson NASA HQ Building, 
300 Hidden Figures Way SW, Washington, DC 20546, USA}

\author{D. Larson}
\affiliation{University of California at Berkeley, Berkeley, CA 94720, USA}

\author{R. Livi}
\affiliation{University of California at Berkeley, Berkeley, CA 94720, USA}

\author{T. Lim}
\affiliation{University of New Hampshire, Durham, NH, 03824, USA}

\author{R. A. Leske}
\affiliation{California Institute of Technology, Pasadena, CA 91125, USA}

\author{O. Malandraki} 
\affiliation{Institute for Astronomy,
  Astrophysics, Space Applications and Remote Sensing of the National
  Observatory of Athens, Vas. Pavlou and I. Metaxa, 15236 Penteli,
  Greece}


\author{D. Malaspina} 
\affiliation{Laboratory for Atmospheric and
  Space Physics, University of Colorado, Boulder, CO 80303, USA}

\author[0000-0001-7224-6024]{W. H. Matthaeus}
\affiliation{Department of Physics and Astronomy, University of Delaware \\ Newark, DE 19716, USA}

\author{D. J. McComas} 
\affiliation{Department of Astrophysical Sciences, Princeton University, Princeton, NJ, 08544, USA}

\author{R. L. McNutt Jr.}
\affiliation{Applied Physics Laboratory, Laurel, MD 20723, USA}

\author{R. A. Mewaldt}
\affiliation{California Institute of Technology, Pasadena, CA 91125, USA}

\author{D. G. Mitchell}
\affiliation{Applied Physics Laboratory, Laurel, MD 20723, USA}

\author{J. T. Niehof}
\affiliation{University of New Hampshire, Durham, NH, 03824, USA}

\author{M. Pulupa}
\affiliation{University of California at Berkeley, Berkeley, CA 94720, USA}

\author[0000-0003-4168-590X]{Francesco Pecora}
\affiliation{Department of Physics and Astronomy, University of Delaware \\ Newark, DE 19716, USA}

\author{J. S. Rankin}
\affiliation{Department of Astrophysical Sciences, Princeton University, Princeton, NJ, 08544, USA}

\author{C. Smith}
\affiliation{University of New Hampshire, Durham, NH, 03824, USA}

\author{E. C. Stone}
\affiliation{California Institute of Technology, Pasadena, CA, 91125, USA}

\author{J. R. Szalay}
\affiliation{Department of Astrophysical Sciences, Princeton University, Princeton, NJ, 08544, USA}

\author{A. Vourlidas}
\affiliation{Applied Physics Laboratory, Laurel, MD 20723, USA }

\author{M. E. Wiedenbeck}
\affiliation{California Institute of Technology, Pasadena, CA 91125, USA}

\author{P. Whittlesey}
\affiliation{University of California at Berkeley, Berkeley, CA 94720, USA}

\begin{abstract}
We present an event observed by Parker Solar Probe at $\sim$0.2 au on March 2, 2022 in which imaging and \emph{in situ} measurements coincide.  During this event, PSP passed through structures on the flank of a streamer blowout CME including an isolated flux tube in front of the CME,  a turbulent sheath, and the CME itself.  Imaging observations and \emph{in situ} helicity and principal variance signatures consistently show the presence of flux ropes internal to the CME.    
 In both the sheath, and the CME interval, the distributions are more isotropic, the spectra are softer, and the abundance ratios of Fe/O and He/H are lower than those in the isolated flux tube, and yet elevated relative to typical plasma and SEP abundances. These  signatures in the sheath and the CME indicate that both flare  populations and those from the plasma are accelerated to form the observed energetic particle enhancements. In contrast, the isolated flux tube shows large streaming, hard spectra and large Fe/O and He/H ratios,  indicating flare sources.  Energetic particle fluxes are most enhanced within the CME interval from suprathermal  through  energetic particle energies ($\sim$ keV to $>10$ MeV), indicating particle acceleration, and confinement local to the closed magnetic structure.  
The flux-rope morphology of the CME helps to enable local modulation and trapping of energetic particles, particularly along helicity channels and other  plasma boundaries. Thus, the CME acts to  build-up  energetic particle populations, allowing them to be fed into subsequent higher energy particle acceleration throughout the inner heliosphere where a compression or shock forms on the CME front.
\end{abstract}


\section{Introduction} 

The NASA Parker Solar Probe (PSP) Mission \cite[]{Fox:2016} is the first to directly explore the environment near the Sun by going there. The Integrated Science Investigation of the Sun (\ISOIS) instrument suite \cite[]{McComas:2016} on PSP provides the first measurements of solar energetic particles (SEPs) in this environment close the Sun using the EPI-Hi and EPI-Lo instruments over the range 0.02–200 MeV/nucleon. \cite{McComas:2016} asked the following questions that are central to PSP science:
\begin{enumerate}
\item	What is the origin of the seed population for solar energetic particles (SEPs)?
\item	How are these SEPs and other particle populations accelerated?
\item	What mechanisms are responsible for transporting the different particle populations into the heliosphere?
\end{enumerate}
Here, we examine the sources of  energetic particles and their seed populations within the flank of a coronal mass ejection (CME) observed by \ISOIS~ on March 2, 2022. The observed particle populations respond to the magnetic fields observed locally at PSP by the Electromagnetic Fields Investigation \cite[FIELDS,][]{Bale:2016} and the solar wind plasma observed by the Solar Wind Electrons Alphas and Protons Investigation \cite[SWEAP,][]{Kasper:2016}. The evolution of the CME is observed globally  by the Wide Field Imager for Solar Probe Plus (WISPR) \cite[]{Vourlidas:2016}.

The reviews by \cite{Gosling:1993} and \cite{Gosling:1994} showed  that CMEs are fundamental in creating the non-recurrent disturbances that disrupt the magnetosphere and drive geomagnetic storms. CMEs accompany large  $\mathbf{J}\times \mathbf{B}$ forces within magnetic flux tubes driven out of their quasi-equilibrium configurations in the corona during CME eruption and acceleration. The  $\mathbf{J}\times \mathbf{B}$ forces cause  acceleration of the plasma, and this acceleration drives the development of  compression regions and  traveling shocks through the heliosphere. The shocks are  known sites of rapid particle acceleration \cite[e.g.][]{Li:2009, Schwadron:2015SEP}.  Large and fast CMEs drive strong shocks that accelerate  high energy particles with large fluences, creating significant radiation risks   \cite[e.g.,][]{Schwadron:2010a, Schwadron:2014b} for astronauts and widespread problems for spacecraft and the instruments and electronics they carry. 

Particle energization begins low in the corona during the fast expansion of a CME.  
The rapid particle acceleration is associated with  the development of first compression regions and then shocks around the emerging CME \cite[]{Gorby:2012, Linker:2014, Schwadron:2014bb}. 
Energization in compression regions \cite[see, ][]{Giacalone:2002, Jokipii:2003} and at shocks
 \cite[e.g.,][]{Fermi:1949, Drury:1983} is caused by the diffusive movement of charged particles  
 across these structures with sharp velocity gradients. In the plasma frame,  charged particles gain energy after 
 each crossing of the velocity gradient, and the total  energy gained depends on the number of  crossings by a particle \cite[]{Bell:1978a, Bell:1978b}. 
 
 Diffusive processes in plasmas are tied to the interactions 
 between charged particles and waves or turbulence within the plasma. Near a shock or any large velocity gradient 
 in the plasma, charged particles encounter fluctuations in the magnetic field, often characteristically Alfv\'enic, that cause 
 changes in the particle pitch-angles. After many such wave-particle interactions, charged particles 
 are  scattered in pitch-angle, which leads to strong diffusion. 
 
 Rapid diffusion at the shock requires short scattering 
 mean free paths \cite[]{Lee:1981, Lee:1983, Lee:2005} and/or a quasi-perpendicular magnetic field for 
 high rates of particle energization  \cite[]{Jokipii:1982, Jokipii:1986, Jokipii:1987}. 
 The strong draping of magnetic fields  about the CME forms
 quasi-perpendicular field structures \cite[]{Schwadron:2015SEP}, and the large anisotropies near the shock or velocity gradient is a known source for wave amplification \cite[]{Li:2009b, Li:2012}. Together, these conditions cause the 
 high traversal rate of the speed gradient, and are therefore associated with rapid particle energization. 

An important feature of the structures that accelerate particles from the low corona is the buildup of the quasi-perpendicular field compression at the front of the CME expansion. The fact that this sheath is draped by magnetic fields containing the plasma that is swept up by the CME suggests that this region should be effective at storing particle populations. Therefore, these sheaths may naturally build up seed populations, and in particular, those released by earlier flaring activity in the vicinity of the eruption.  Subsequent acceleration near the shock or compression  thereby enhances the suprathermal and energetic particle populations  within the sheath. Observations by the Integrated Science Investigation of the Sun \cite[\ISOIS , ][]{McComas:2019}  show precisely this process. The draped fields in front of a CME build up the energetic populations of charged particles from the surrounding plasma \cite[]{Schwadron:2020seeds}. 


Pre-existing seed populations were hypothesized as the result of flaring and/or nanoflaring  at the Sun \cite[]{Parker:1988}. The presence of enhanced $^3$He throughout observed events provides direct evidence that flares contribute to energetic particle seed populations  \cite[]{Mason86, Mason:2002, Reames:1999, Desai:2003}. The generation of seed populations from flares, the subsequent compressive build-up (pre-conditioning) of these populations, and then the higher energy acceleration as strong compressions and shocks form further out in the heliosphere supports the unifying role of the CME. 

\begin{figure}[ht]
\centering
\includegraphics[width=0.9\columnwidth]{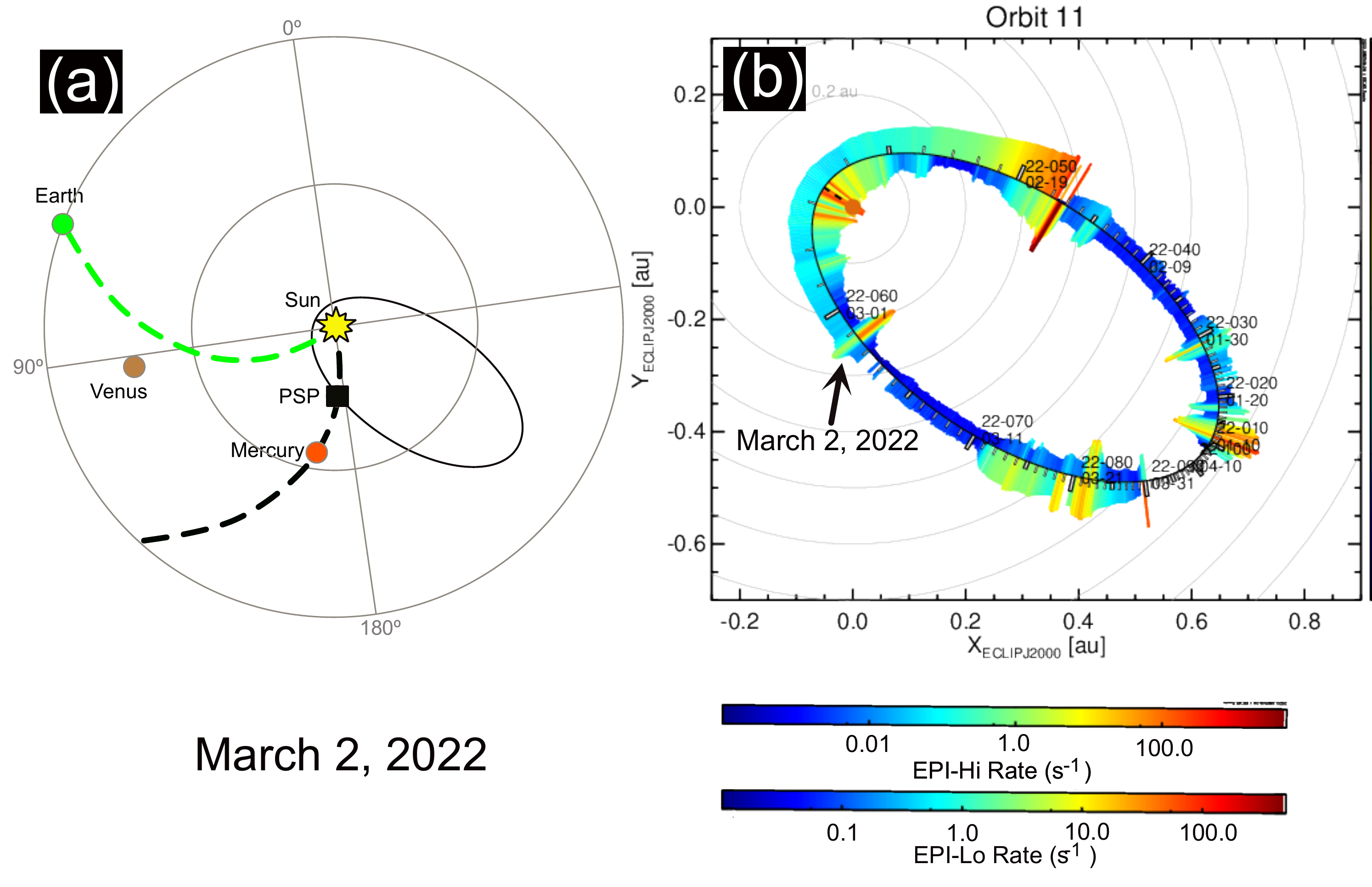} 
\caption{(Panel a) The location of the PSP spacecraft, Earth, Venus and Mercury, Parker spirals on the PSP- and Earth-connected field lines on March 2, 2022, during the event reported here. We also include PSP’s orbit 11, and Carrington Longitudes at 0$^\circ$, 90$^\circ$, and 180$^\circ$. (Panel b) Summary of the count rates of EPI-Lo (interior to orbit), and EPI-Hi (exterior to orbit).  EPI-Lo count rates are triple coincidence (time-of-flight and energy) protons, 67-206 keV, summed over all 80 apertures of the instrument for approximately 2$\pi$ sr coverage.  EPI-Hi count rates are protons from the LET1 telescope A side (pointing up the nominal Parker Spiral, with 45$^\circ$ half-width FOV), range 1, which is responsive to incident energies 0.71-2.8 MeV. }
\label{fig:f1}
\end{figure}

On the March 2, 2022 during Orbit 11 of PSP, we were provided with a rare opportunity to observe the development of the seed population within structures surrounding a small CME. 
Figure \ref{fig:f1} shows an overview of the particle events observed in PSP Orbit 11. A series of events are observed throughout the orbit, but we focus here on the March 2, 2022 event when PSP was overtaken by the flank of a CME. The event occurred as the PSP spacecraft moved out of perihelion, and data for the event was recorded while PSP was near relatively close to the Sun, at $\sim 0.2$ au. 

During the Orbit 11 perihelion passage, EPI-Lo observed increased count rates associated with relatively low energy ($<100$ keV/nuc) ions and relatively soft energy spectra (differential flux scales as $E^{-4}$ to $E^{-6}$ where $E$ is energy) near the current sheet that stream away from the Sun. The ions observed near the Orbit 11 encounter are characteristically similar to those observed by \ISOIS~ on previous current sheet crossings \cite[]{Desai:2022} during encounters 7, 8 and 9. \ISOIS~ observations during the orbit 11 encounter show H, He, O, and Fe  in  the energy range from below 30  to $\sim 100$ keV/nucleon.  The association with current-sheet crossings traditionally implicates magnetic reconnection-driven processes as the most-likely source of particle energization. 

Recent work \cite[]{Mitchell:2020}  studied a solar energetic particle event observed by \ISOIS~ during PSP's first orbit (day-of-year 314-316, 2018)  associated with the release of a slow coronal mass ejection close to the Sun ($\sim 0.23$ au). This study likened the event with the energetic particle population produced in high current density structures associated with auroral phenomena in planetary magnetospheres. In this ``auroral pressure cooker'' mechanism, an electric field creates strong parallel currents that amplify a broadband distribution of waves. These waves, in turn,  accelerate ions over a large range of charge/mass ratios.  \cite{Temerin:1992} compared auroral plasmas with flare-associated plasmas, concluding that ion heating from resonances with electromagnetic ion cyclotron (EMIC) waves  should energize particles efficiently in the Sun’s corona. The precise mechanism energizing ions near  PSP current-sheet crossings close to solar encounters remains an area of active investigation. Studying and understanding the processes that accelerate ions away from strong compressions or shocks may be critical for developing  more generalized views of the processes responsible for energetic ions near the Sun and, by extension, to other heliospheric and astrophysical plasma environments. 

Our study is motivated by its association with a slow-moving CME that was initiated at the Sun on March 1, 2022. Without clear signatures of a shock or a strong compression region, we could conclude the absence of energetic particles. We show observations indicating the opposite.  \ISOIS~ observations near the Sun indicate the local sources of seed populations that contribute to the build-up of energetic particles in the flank of the CME, and indicates that CMEs act as reservoirs for seed populations.  The paper is organized as follows. In \S 2, we provide detailed analysis of imaging observations and \emph{in situ} data from PSP  yielding global context for the structures observed in the period from March 1 to 3, 2022. We  assess the energetic particle composition and anisotropies in  \S 3. In \S 4, we discuss the diagnostics of turbulence over the periods studied,  and in \S 5 we analyze and fit observed energetic particle spectra.  We discuss key results in \S 6,  and present our  conclusions in \S 7. An appendix is provided that discusses the energization of ions from flux tube disruption. 

\section{PSP Passes into the Flank of the March 2, 2022 Streamer-Blowout CME: Global Context from Imaging and \emph{In situ} Observations}

The  March 2, 2022 event is a streamer-blowout (SBO) CME that became apparent in the coronagraphic fields-of-view (FOV) on March 1 at approximately 00:35:53 UTC and crossed the WISPR FOV between March 1, 12:00 UT to March 2 $\sim$17:00 UT. The latter interval coincides with the \ISOIS~ and SWEAP measurements of transient signatures   and mark this as one of the rare events where both imaging and \emph{in situ} measurements coincide.  

\subsection{Imaging Observations from Feb 28 –  March 3, 2022}

\begin{figure}
\centering
\includegraphics[width=\columnwidth]{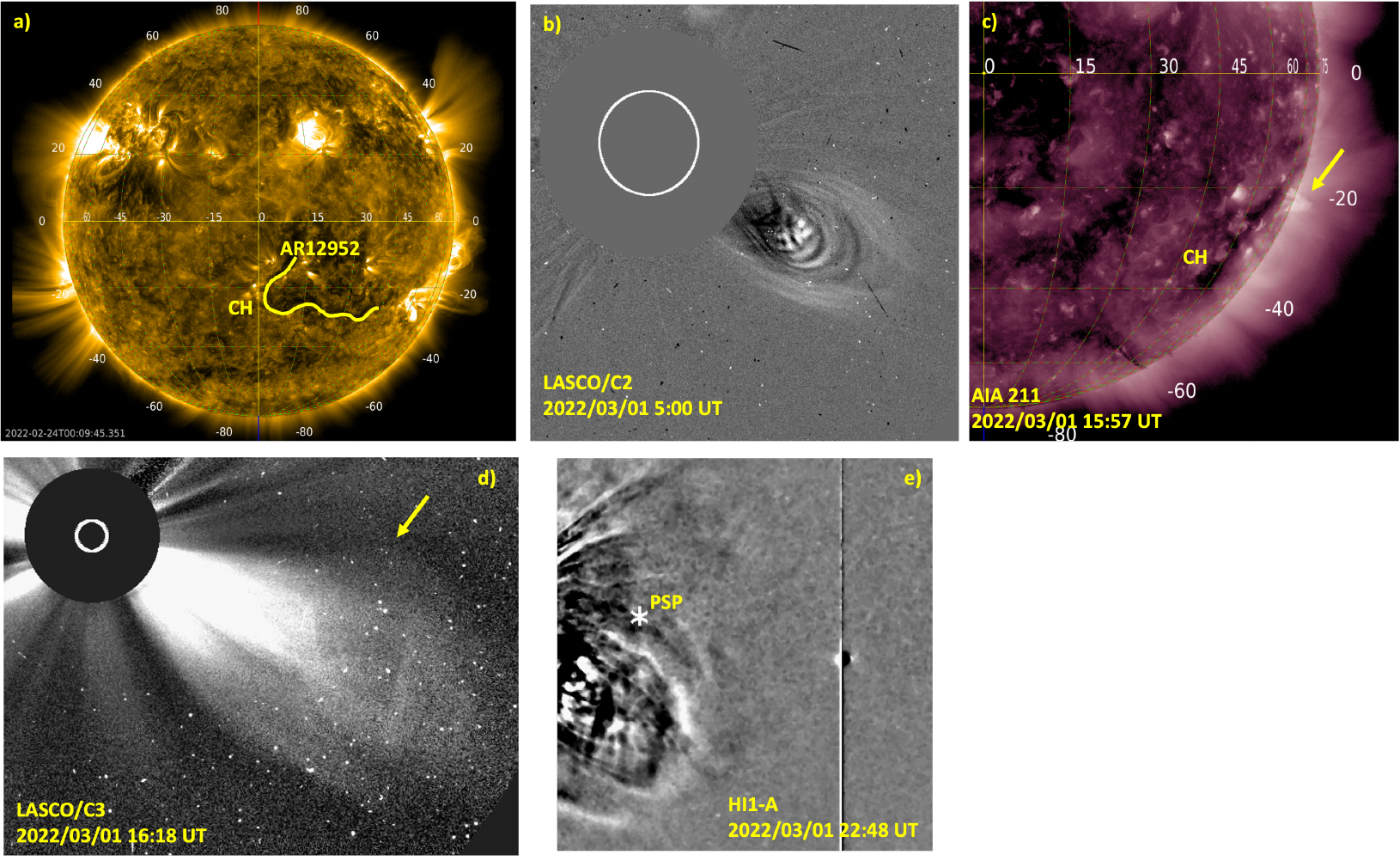}
\caption{Summary of 1 au imaging observations of the CME. a)  The candidate source region is marked with the dashed line. The westernmost edge of the CME likely lies along AR12952 due to the presence of a coronal hole (‘CH’). B) The running difference C2 image shows a slow-rising flux-rope type CME pushing the overlying streamer to its flanks. c) The best low corona evidence of a post-CME loop system is seen in AIA 211 images marked by the arrow. d) The CME extends considerably along the north (marked by arrow) and e) intercepts the PSP location (star) even though the bulk of the CME moves below the PSP orbit plane. }
\label{fig:f2z}
\end{figure}

\begin{figure}
\centering
\includegraphics[width=0.8\columnwidth]{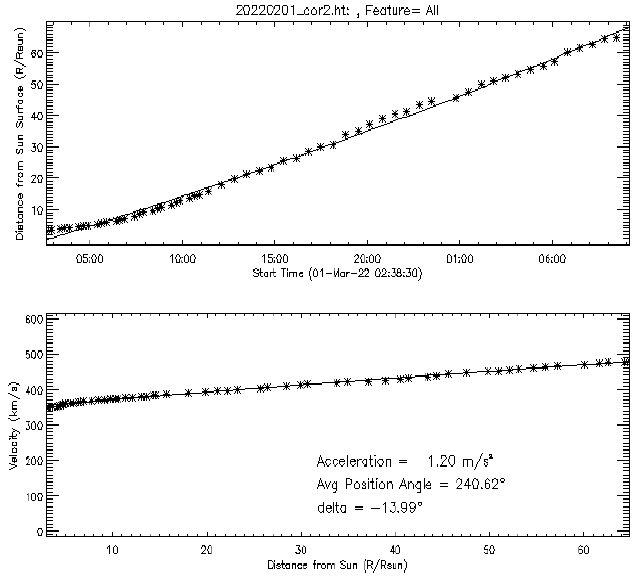}
\caption{Kinematic of the CME leading edge measured in the COR2 and HI1 FOVs. The CME accelerates slowly reaching a speed of 440 km/s at the approximate location of PSP ($\sim 47$ R$_s$). }
\label{fig:f2y}
\end{figure}

A gradual streamer expansion, marking the early stages of an SBO-CME \cite[]{Vourlidas:2018} became apparent in the LASCO/C2 FOV from Feb 28, 22:36 UT onwards. A rather diffuse front emerged at around 00:35 UT on March 1. The full CME, exhibiting a clear flux-rope morphology with multiple striated features \cite[]{Vourlidas:2017} was visible by 05:00 UT (Fig. \ref{fig:f2z}). We tracked the CME front in the STA COR2 and HI-1 FOVs up to about 65 R$_s$. The kinematic analysis shows a gradually accelerating CME (Fig. \ref{fig:f2y}) reaching speeds of $\sim$500 km/s by 65 R$_s$. 

As is common for most SBO-CMEs, identification of the source region is difficult. In this case, the task is complicated by the source being largely behind the western solar limb. Based on an extensive review of the solar activity from Feb. 21 to March 2, the most likely source lies between and including, AR12952 and the region westwards. The dashed yellow line in Figure \ref{fig:f2z}a marks the candidate polarity inversion line (PIL) on February 23, when the region is well-observed by Earth. AR12952 exhibited a couple of CMEs along the North-South PIL crossing the region in the days before the event. The March 1$^\mathrm{st}$ CME could arise from anywhere along the PIL that extends 30° in the north-south and 45° in the east-west direction. There are no indications of a large-scale liftoff of the filament overlying the PIL during the overlying streamer swelling on February 28 through March 1. However, SDO/AIA observations in the 211 channel (Fig. \ref{fig:f2z}c) show a short bright arcade associated with a small-scale ejection at the tail-end of the CME. The signature is consistent with post-eruption arcades and aligns with a flux-rope signature in the WISPR images that will be discussed later in this section. The AIA and EUVI observations show faint loop expansion over the south-west limb consistent with an expanding far-side CME. In summary, the EUV observations indicate a slow far-side expansion, most likely along the PIL marked in Figure \ref{fig:f2z}  with its easternmost boundary lying along the north-south PIL crossing AR12952. 

The coronal and inner heliospheric observations are consistent with the EUV signatures. We focus on the aspects relevant to the particle observations here. Although the CME front is moving in a southern direction, approximately 30° south, the overall CME envelope is expanding northwards as it can be gleaned by the LASCO C2 and C3 images (Fig. \ref{fig:f2z}b, d). The northern extension (marked by the yellow arrow in Fig. \ref{fig:f2z}d) intercepts the PSP position (Fig. \ref{fig:f2z}e). PSP is located 108° west of Earth and only 17° west of AR12952. Since the CME originates westward of AR12952 and has a width of at least 40°, given the very clear flux-rope morphology and its appearance on the WISPR FOV, PSP is well within the longitudinal envelope of the CME and we expect to see signatures  \emph{in situ}.

\begin{figure}
\centering
\includegraphics[width=0.6\columnwidth]{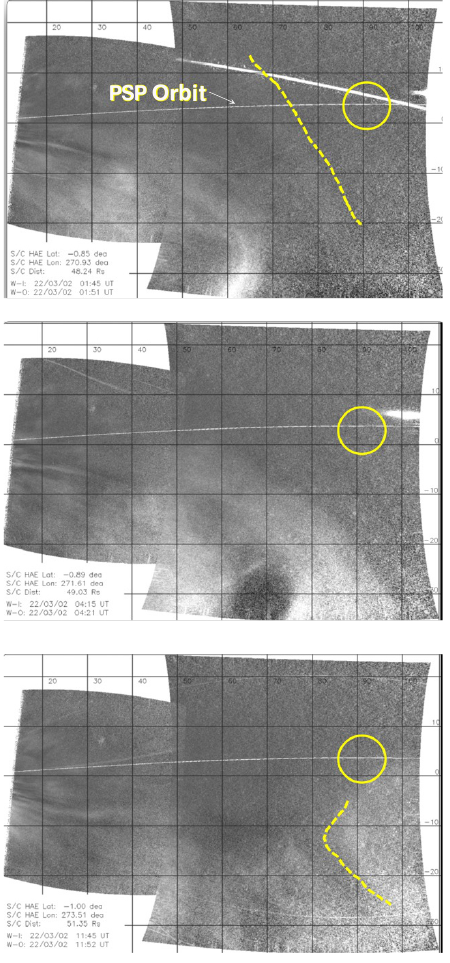}
\caption{WISPR observations of the March 1st CME. The frames, taken from the accompanying movie, show the incoming CME front (top, yellow dashed line), the encounter time (middle), based on the images, and the end of the interaction (bottom) marked by the crossing of a V-shaped feature (dashed line) PSP is moving along the white line, marked in each panel, and towards the yellow circle (approximately 90° elongation). }
\label{fig:f2x}
\end{figure}

 `Local' imaging of the CME is provided by WISPR. Figure \ref{fig:f2x}  shows three snapshots of the WISPR composite movie that accompanies the article. The images have been processed to enhance diffuse structures in an effort to make visible the CME front, most relevant to the particle observations. The structures can be discerned much easier in the movie but the yellow line in the top panel marks a front corresponding to the northern CME flank seen in the C3 and HI1 images (Fig. \ref{fig:f2z}c, e, respectively). The front becomes progressively more diffuse with time. This is the expected behavior for a structure approaching the spacecraft. PSP is moving along the white line, marked in each panel, and towards the yellow circle (approximately 90° elongation). Based on the images alone, we estimate that the PSP-CME interactions starts on March 2$^{\mathrm{nd}}$ at around 04:15 UT and last until the passage of a V-shaped structure, at around 12:15 UT. This structure is highly reminiscent of the V-shaped tail-end of flux-rope CMEs and aligns approximately with the location of the post-CME loop system detected in AIA 211 (Fig. \ref{fig:f2z}c).
 The fact that both the northern flank and the flux-rope entrained in the CME are visible in WISPR further reinforces our  assessment that the CME is wide enough to engulf the PSP position.

\subsection{Global Context from SEPs and In Situ Observations from 2022 March 1 – 3}

\begin{figure}
\centering
\includegraphics[width=\columnwidth]{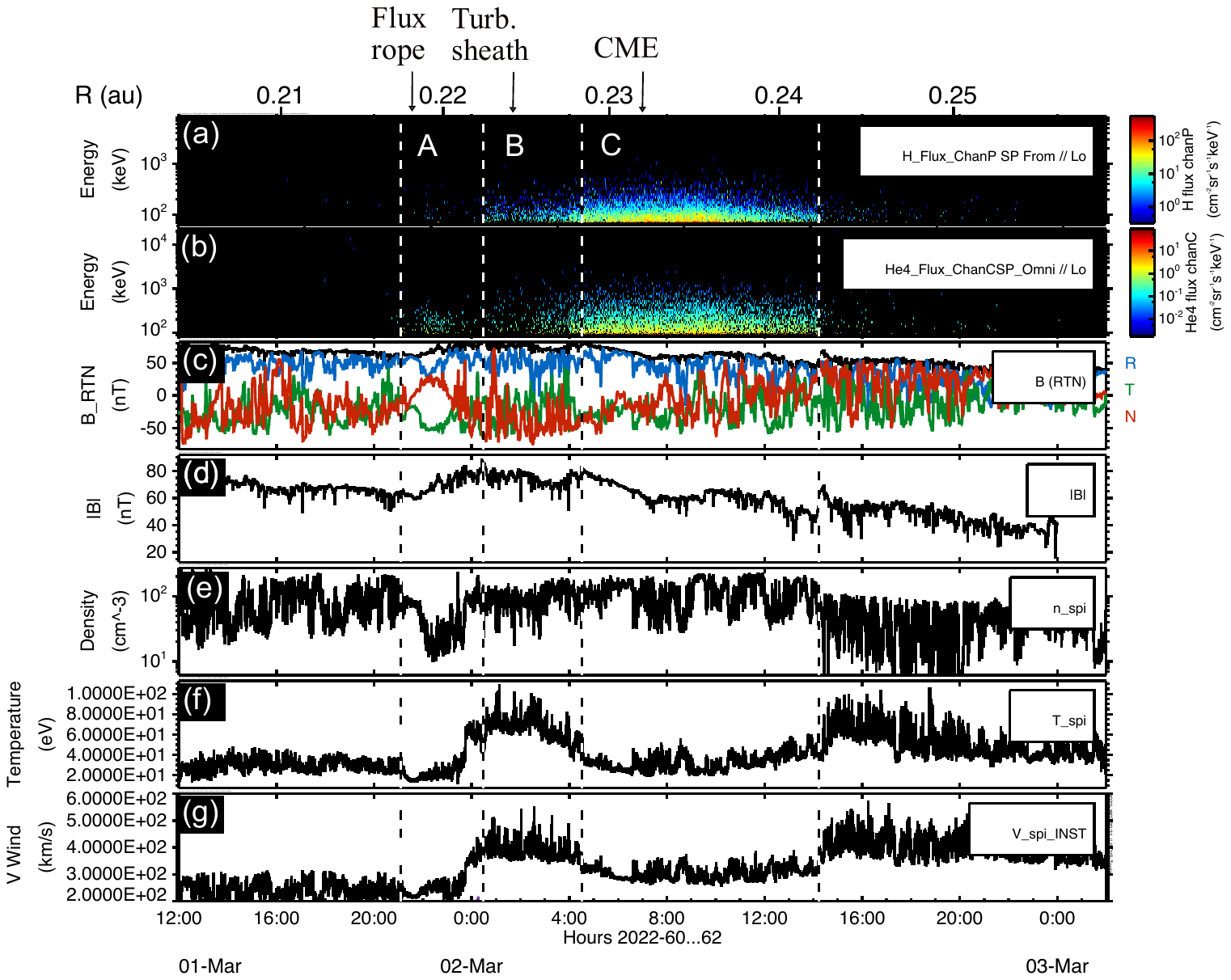}
\caption{The EPI-Lo differential fluxes of protons (a) and $^4$He ions (b), the FIELDS magnetic field (c), the magnetic field strength (d), and SWEAP solar wind density (e), core plasma temperature (f), and solar wind speed (g). In the text, we identify three separate intervals: interval A is associated with an isolated flux tube; interval B is a turbulent sheath in front of the CME ejecta, and interval C exists within the CME itself. We observe significant fluxes of ions throughout the event and in each of the intervals identified, but the character of these populations, including their compositions, changes substantially in each of the intervals identified. }
\label{fig:f2}
\end{figure}

 Figure \ref{fig:f2} shows the energetic particles fluxes of protons and $^4$He measured by EPI-Lo together with FIELDS magnetic field, and SWEAP solar wind density, temperature and solar wind speed. Interval A (between 2022-060T21:05:27.270 and 2022-060T23:44:08.319 UTC) shows a rotation in the magnetic field and a small build-up in magnetic field strength, a drop in temperature and a relatively slow solar wind speed. Appendix A.1 considers the force balance of the observed flux tube, which is not in equilibrium. The $\mathbf{J}\times \mathbf{B}$ forces of the flux tube drive its expansion as it is overtaken by structures associated with the coronal mass-ejection behind it. This outward expansion of the flux rope is consistent with the lower plasma pressure within the structure.

The lower temperature internal to the CME  indicates expansion of the plasma, and the successive rotations in the magnetic field  are  associated with flux tubes within the internal CME structure. The orientation of the magnetic field is predominantly radial and together with PSP’s location on the CME flank shows that the CME swept over spacecraft. \cite{McComas:2023} show separate observations of a CME leg where energetic particles populations are almost entirely absent. In contrast, we observe enhanced energetic particle signatures, showing that energetic particles are entrained within the magnetic structures of the CME. 

\begin{figure}
\centering
\includegraphics[width=\columnwidth]{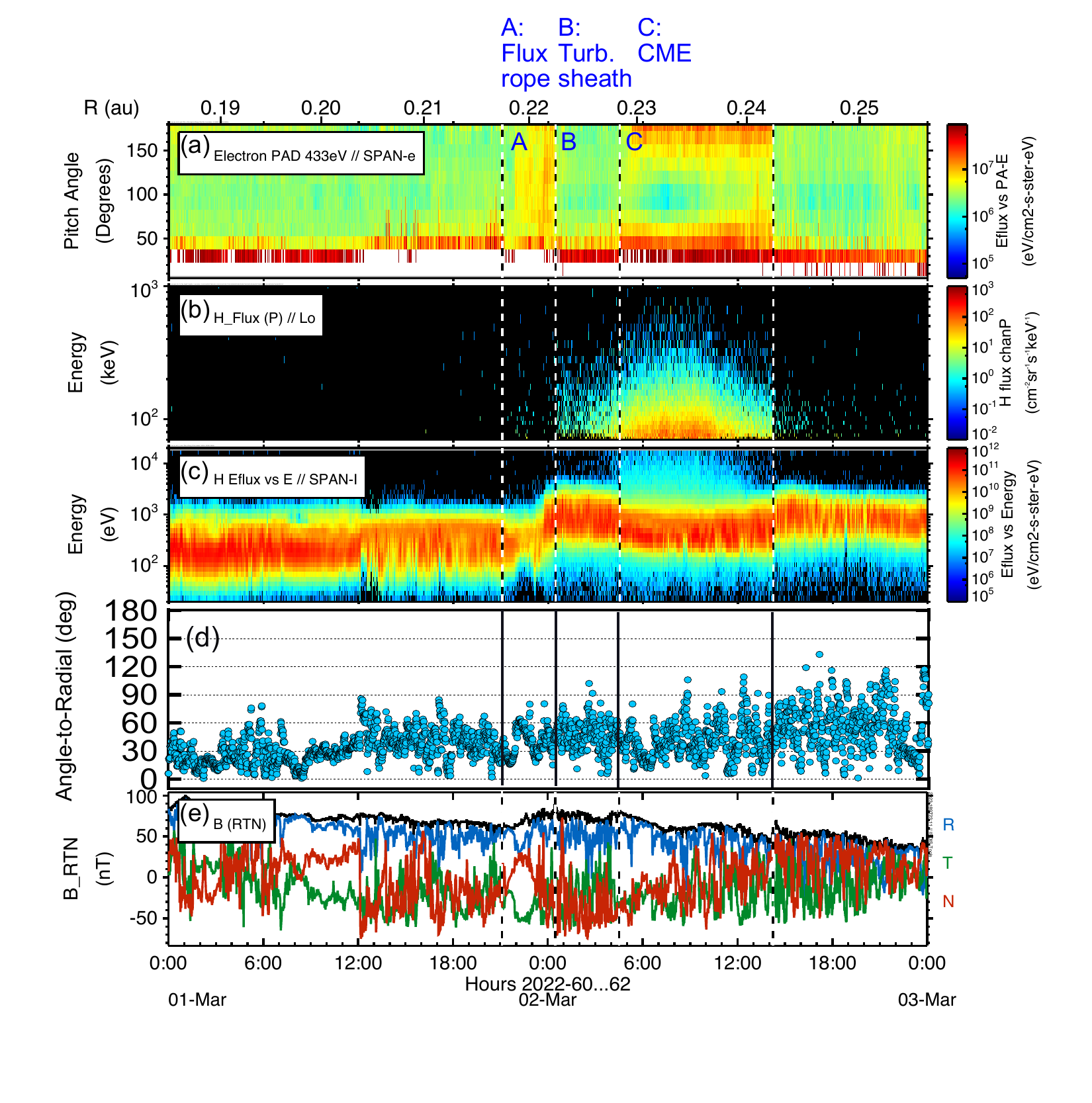}
\caption{(Top) The PSP/SWEAP/SPAN-e electron pitch-angle observations show bi-directional streams in intervals A and C. Shown from top to bottom are the electron pitch-angle distribution (a), the \ISOIS~/EPI-Lo differential proton flux (b), the PSP/SWEAP/SPAN-I energy flux (c), and the PSP/FIELDS magnetic field measurements (d). The dashed vertical lines indicate the plasma interval boundaries identified in Figure \ref{fig:f2} corresponding to the isolated flux rope (interval A), the turbulent sheath (interval B), and the CME  (interval C).  }
\label{fig:f3}
\end{figure}

Figure \ref{fig:f3} shows the electron observations during the event observed by the SWEAP instrument. Intervals A and C show the classic signatures of counter-streaming electrons associated with magnetic structures tied at both ends to the Sun. We also note in panel (b) the large energy flux of electrons within the flux rope (Interval A), which is consistent with a strongly elevated current. Within Interval B, we see the loss of counter streaming, showing that the region is comprised by simply connected open magnetic fields, and is consistent with the identification of the sheath compressed by the CME  within Interval C. Note the strong reduction in electron flux within the flux rope. This is consistent with the structure having been displaced by the sheath, and the expansion of the flux tube in the presence of a strong internal stress.

\begin{figure}
\centering
\includegraphics[width=0.8\columnwidth]{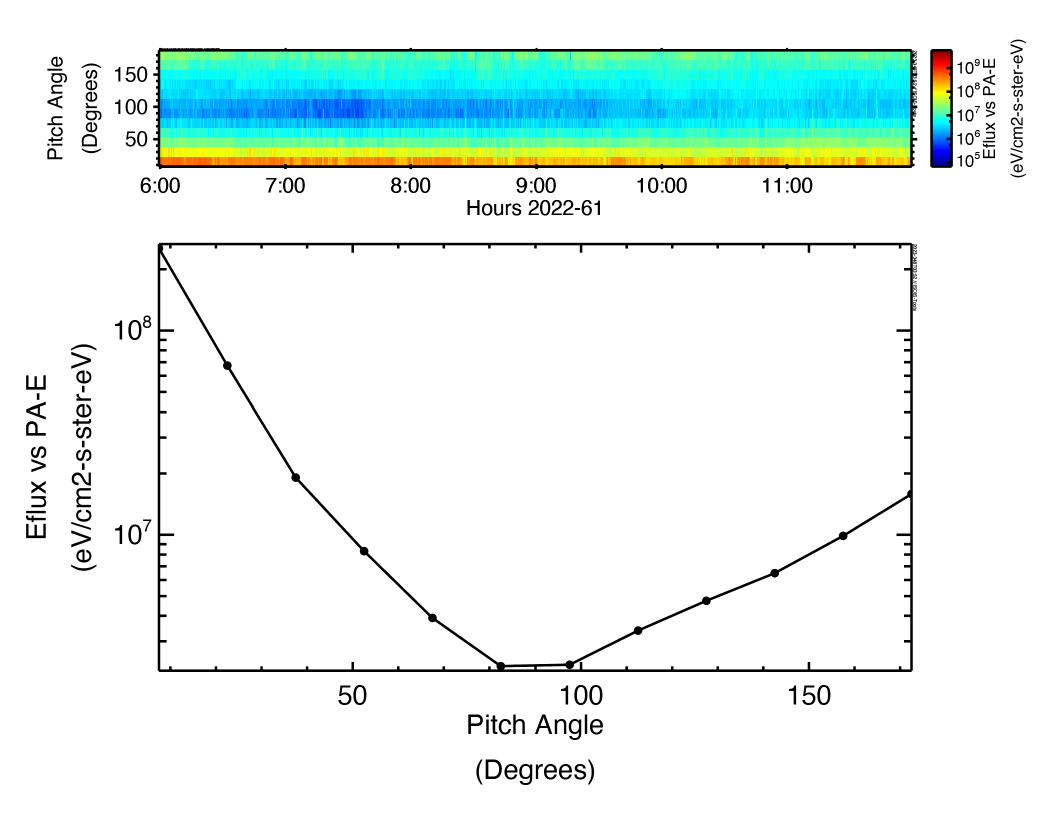}
\caption{(Top) The PSP/SWEAP/SPAN-e electron pitch-angle observations showing bi-directional streams in interval C, (Bottom) and the electron energy flux distribution as a function of pitch-angle.  The counter-streaming distribution is evident, but highly asymmetric consistent with a field connection closer to the Sun on the near-radial outward polarity flank of the CME.  }
\label{fig:f3a}
\end{figure}

The counter-streaming electrons observed within Interval C show that the structure is closed magnetically. The large asymmetry (see Figure \ref{fig:f3a} for the electron energy flux distribution) in counter-steaming shows the vast predominance of electrons streaming outward from the Sun and a significant but smaller stream of electrons with inward streaming back to the Sun. \cite{Phillips:1992}  studied counter-streaming electron fluxes from 39 CMEs observed from ISEE-3. Generally, dominant outward electron asymmetries (with more outward  than inward electrons) were observed 75\% of the time.   The most pronounced asymmetries with $\sim$ 4 times the number of outward vs inward electrons were associated with  near-radial magnetic fields. In the March 2, 2022 event detailed here, the asymmetry observed has an average of 8 times the energy flux in the outward versus the inward hemisphere, consistent with the large asymmetries observed by \cite{Phillips:1992} in near radial magnetic field structures. The large asymmetry in counter-streaming is consistent with a field connection much closer to the Sun on the outward polarity flank of the closed field structure. 

To investigate the presence of helical structures and sharp boundaries, we employ two measures that have been previously applied to the solar wind and simulation data. Helical features in magnetic field measurements can be revealed using the magnetic helicity technique developed in \citet{pecora2021identification}, based on the theory first developed by \citet{matthaeus1982measurement}. Following their prescription, the magnetic helicity at a certain scale $\ell$ can be estimated using the non-diagonal terms of the magnetic field autocorrelation tensor, $R_{ij}(r)=\langle B_i(x+r)B_j(x) \rangle$, as $ H_m(x,\ell) = - \int_0^\ell dr_i \; \epsilon_{ijk} R_{jk} $, where $\langle \dots \rangle$ indicates an average over a suitable interval, and $r$ are the increments along the direction $i$. For spacecraft measurements, the increments are intended to be taken in the time domain and could be converted into spatial distances using the Taylor hypothesis \citep{taylor1938spectrum}. Usually, the relative motion of the solar wind with respect to PSP is mostly radial, therefore, the direction of the increments can be considered to be along the coordinate R of the RTN (Radial-Tangential-Normal) reference frame. The two transverse directions $j$ and $k$ are associated with the T and N coordinates \citep{pecora2021parker}.

A convenient measure of magnetic helicity is relative to the magnetic field fluctuations. 
It is possible to estimate 
a characteristic value for a normalized version of the helicity, $\tilde{H}_m = \frac{H_m}{\lambda E} $, where $\lambda$ is a characteristic length and $E$ is the magnetic energy. 
The values of $\tilde{H}_m$ whose magnitude is larger than such a threshold are reasonably due to helical structures rather than random fluctuations.

The second complementary technique, the partial variance of increments \citep[PVI,][]{greco2008intermittent},  emphasizes the small-scale features of the magnetic field such as discontinuities and current sheets. The PVI is defined as suitably normalized magnetic field vector increments $\Delta \mathbf{B} = \mathbf{B}(t+\tau)-\mathbf{B}(t)$ evaluated at a certain lag $\tau$, namely $\mbox{PVI}(\tau) = \frac{ |\Delta \mathbf{B}| }{\langle |\Delta \mathbf{B}|^2 \rangle}$.  This quantity, as the magnetic helicity, requires a threshold value to distinguish possible coherent structures from statistical noise. Different threshold values usually correspond to different regions of the turbulence fabric \citep{matthaeus2015intermittency}. Generally speaking, $\mbox{PVI} \gtrsim 2.5$ can account for the presence of discontinuities \citep{servidio2011statistical}.   The time lag $\tau$ is tuned to structures of interest. In our analysis, the time lag was tuned to identify the boundaries of the large-scale MHD structure at the scale of several tens of $d_i$, where $d_i$ is the skin depth (i.e., the Alfv\'en speed divided by the proton cyclotron frequency).

\begin{figure}[ht]
    \centering
    \includegraphics[width=\textwidth]{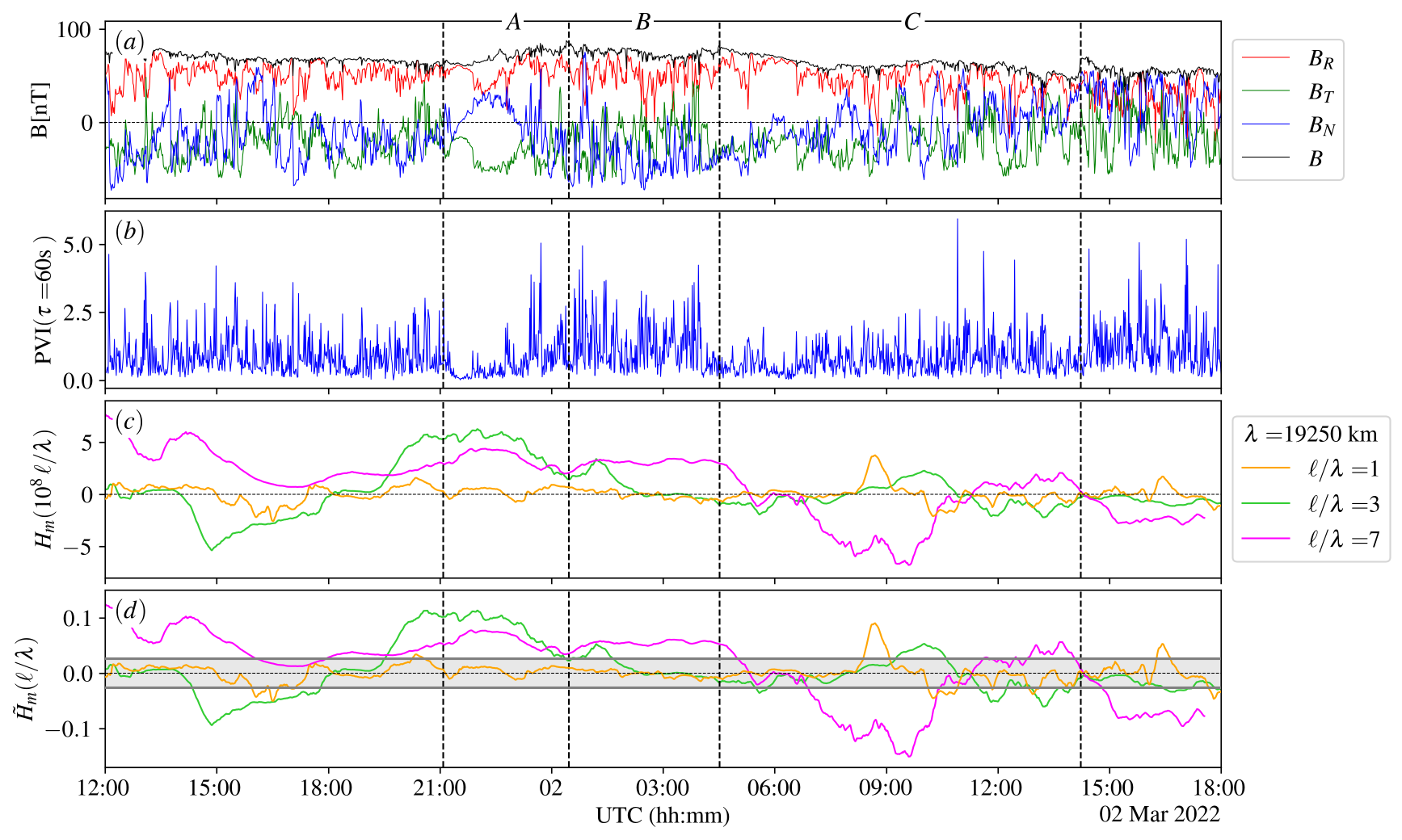}
    \caption{Analysis of magnetic features over 30 hours that include the flux rope position (interval A), the following turbulent sheath (interval B), and the CME  (interval C).
    The magnetic field (a) is reported for convenience.
    The PVI (b) shows a reduction at the position of the flux rope (A) and highly enhanced activity within the sheath, and a somewhat reduced activity in the CME.
    The flux rope helical signature is shown clearly by the magnetic helicity (c) signal at the scale of 3 correlation lengths. Other helical structures at the scale of 7~$\lambda$ appear within the CME, together with a slight bipolar feature at 1~$\lambda$.
    The normalized helicity (d) confirms the presence of such structures where the values of $\tilde{H}_m$ are larger (in magnitude) from the uncertainty region (gray-shaded strip).}
    \label{fig:hmpvi}
\end{figure}

Figure \ref{fig:hmpvi} shows the above-described quantities (PVI, $H_m$, $\tilde{H}_m$) within a time window that includes the intervals of interest. The PVI has been computed with a time lag of 60 seconds. The magnetic helicity has been evaluated at three different scales: 1, 3 and 7 correlation lengths $\lambda$. The magnetic helicity at the smallest scale shows no particular features, except for a bipolar signature at the center of the CME interval (Interval C), as also observed in \cite{McComas:2023}. The intermediate-scale $H_m$ shows a net positive signal where the leading flux rope is (Interval A), in correspondence with the expected reduction of the PVI activity \citep{pecora2021identification}. At the largest scale, the magnetic helicity shows a different behavior. At the leftmost boundary of Figure \ref{fig:hmpvi}, it is noticeable that the helicity is increasing going to the left; this is due to the presence of a very large helical structure between March 1 06:00 - 12:00 that we do not show because it is beyond the scope of this paper. Presumably, the helical signal at this scale that permeates the A and B intervals is only a signature of the declining part of that structure. However, within the CME  (Interval C), there is a negative helicity signature where the PVI signal is low, and its right boundary coincides with a series of very high PVI peaks.

Magnetic helicity is a scale-dependent quantity, and  integrates all the ``information'' up to its scale $\ell$. In Interval C, we find evidence of a large flux rope ($ \ell \sim 7 \lambda $) with a negative sign of helicity that is locally composed of smaller flux ropes with opposite-sign helicity (e.g.,  $\ell \sim $ 1 -  3 $ \lambda $). Note that the scale-dependent changes in the behavior of helicity  is seen commonly in laboratory plasmas  \cite[]{Taylor:1974}. A hierarchy of helical structures is present in Interval C, causing  the magnetic helicity to emerge with the sign associated with the smaller flux ropes at smaller scales, and the sign of the helicity changes when the integration scales include contributions from the larger opposite-signed helical structure.

The association of PVI with energization is well-known observationally,  \cite[e.g.,][]{Osman:2012, Khabarova:2015, Khabarova:2016}. Specifically, large PVI values are locally associated with higher temperatures. The  PVI-T relationship emerges as a statistical property when analyzing large amounts of data.
Such higher temperatures do not necessarily imply that the energization mechanism is happening locally \cite[]{Tessein:2016}.
In our case, the uniformly large PVI values can be associated with a highly turbulent environment (such as the sheath, period B), or with boundaries of flux tubes \cite[as observed in][]{Pecora:2019}. In period C, the flux tubes appear on multiple scales and act as conduits in which energetic particles can flow preferentially (trapping), or the walls of such tubes can keep particles out (exclusion) \cite[]{Pecora:2021}.
The  large-PVI values in period C are found at the boundaries of large (in magnitude) helicity events. Therefore, these flux tube boundaries act as conduits, but not necessarily active acceleration sites. 

\begin{figure}[ht]
    \centering
    \includegraphics[width=\textwidth]{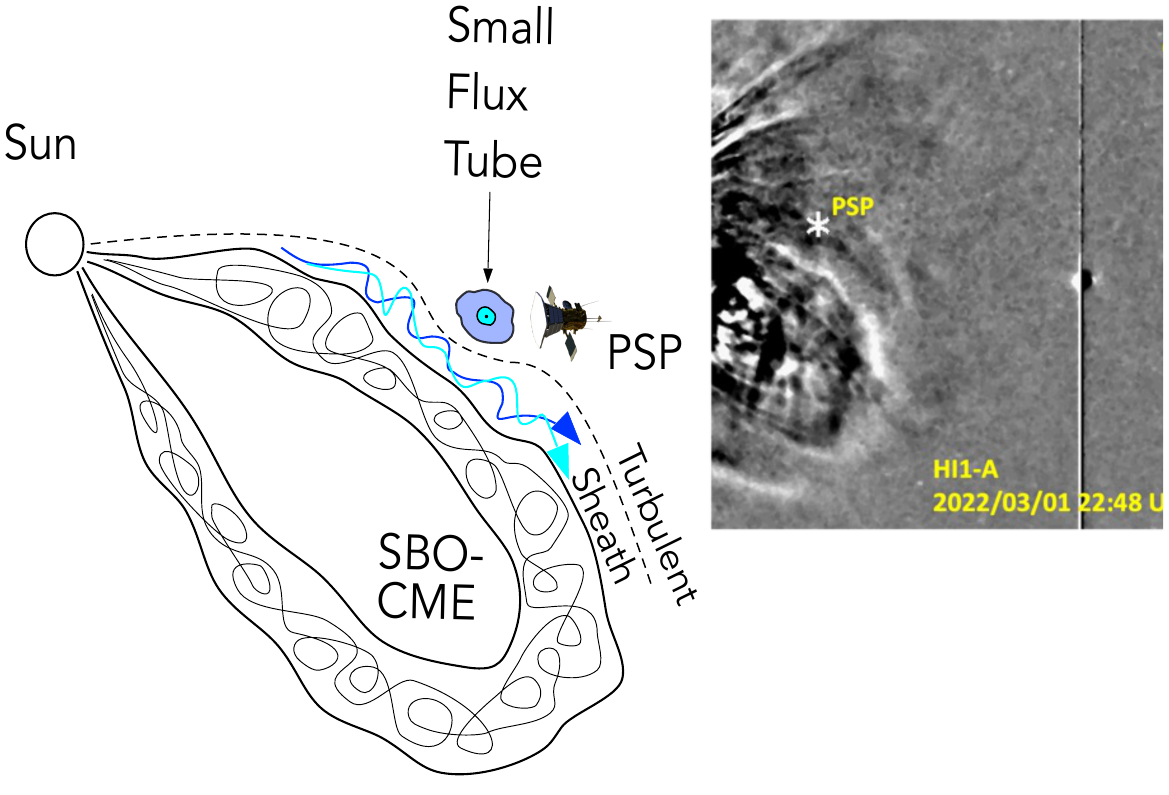}
  \caption{Sketch of the CME, turbulent sheath and displaced flux tube that overtake PSP on March 2, 2022. }
  \label{fig:f5}
  \end{figure}

We sketch  the large-scale configuration observed in Figure \ref{fig:f5}. Note that the magnetic structure of the displaced flux tube and that of the internal structure of the CME  differ substantially. In fact, the field observations suggest multiple flux tubes internal to the CME, and the presence of the sheath, and the small flux tube in front of it provide indications of the interactions between multiple field structures. The magnetic flux tube appears to be an independent structure from the CME. This could indicate that the CME material drives out the small flux tube from its original configuration close to the Sun. 

It is noteable that erupting CMEs  must break through closed flux tubes that work  to hold the CME in place until the outward plasma pressure and Maxwell stresses associated with CME eventually overcome those of the closed flux tubes. The breakout process   accompanies the interaction between the accelerating CME material and closed flux tubes being driven out of their coronal environment, as observed in this case. Disentangling these structures further from the Sun is  complex, and the observations shown here provide an essential view of the interaction between the CME, the surrounding solar wind plasma, and the magnetic flux tube that is lifted out of its original position in the corona.

 \section{Energetic Particle Composition and Anisotropies in the 2022 March 1 - 3 Flux Tube, Sheath and CME}

\begin{figure}[ht]
    \centering
    \includegraphics[width=0.9\textwidth]{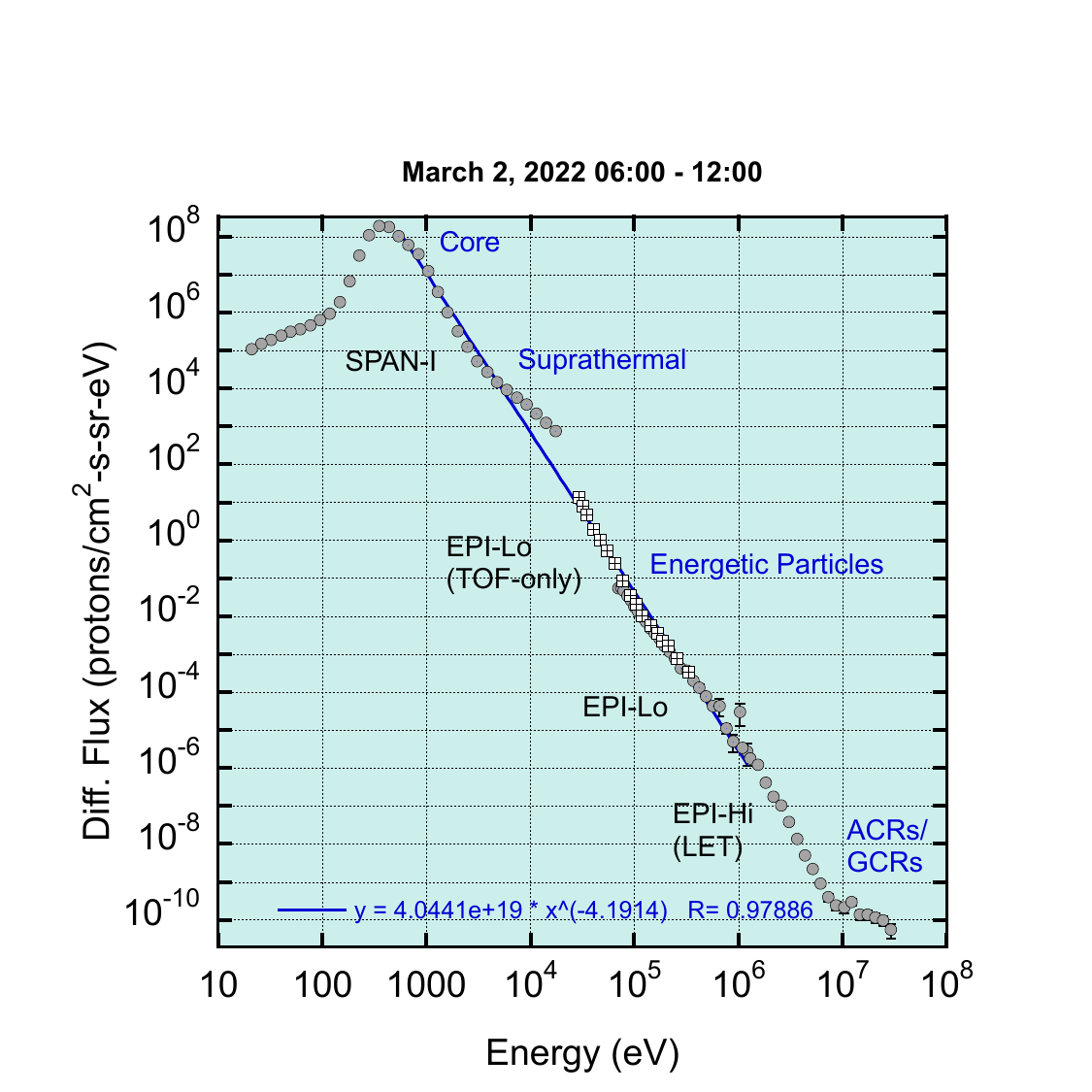}
  \caption{The combined spectrum of protons (Interval C) from the core distribution, through suprathermal, and energetic proton energies during the  March 2, 2022 event (from 06:00 to 12:00 UTC). The suprathermal distribution appears to connect up through energetic particle  distribution, while showing some excess relative to the connected slope  at energies near $\sim 10$ keV. 
We have  included TOF-only ion data  (crossed squares) for all apertures excluding wedge 2 and several additional apertures that are   affected by UV due to holes from dust impacts. TOF-only data is dominated by protons, but does include small contributions from heavy ions typically $<1$ \% at the energies shown.  
  }
  \label{fig:f5a}
  \end{figure}

The distributions observed in Interval C show a  suprathermal population (Figure \ref{fig:f3}, panel b), and the suprathermal flux of particles extends all the way down close to the core of the suprathermal population (\ref{fig:f3}, panel c).  Figure \ref{fig:f5a} shows the connected distribution from suprathermal energies up through energetic particle energies. Notably, we have also included TOF-only ion data  for all apertures excluding wedge 2, which is badly affected by UV due to holes caused by dust impacts.   Further, we have removed several of the apertures adjacent to dust holes that are also affected in the TOF-only data. TOF-only data is dominated by protons at the energies shown but contains small contributions from heavier species (e.g., typically He contributes less than 1\% at around 40 keV in the TOF-only data). While a similar spectral slope connects these populations across more than three decades in energy, we also observe an excess in fluxes in the suprathermal population at $\sim$ 10 keV.  The observed spectral slope $\sim$ -4.2 indicates a soft energy spectrum, consistent with a population of particles that has softened through cooling in the expanding CME. 

Given the relatively small CME, it is surprising to see such strong energetic particle signatures. In this section, we discuss the energetic particle signatures in greater detail, which demonstrate composition  tied to flare populations  in the isolated flux tube (Interval A), but not in the sheath or the CME (Intervals B and C).  Flares typically create large anti-sunward streaming with strong outward anisotropies, large Fe/O ratios ($\sim 1$ or larger) and large He/H ratios ($> 0.01$). The observations in the sheath and particularly in  the CME reveal distributions that are closer to being isotropic with Fe/O ratios and He/H closer to what is expected from acceleration of the the coronal plasma populations. 

\begin{figure}[ht]
    \centering
    \includegraphics[width=\textwidth]{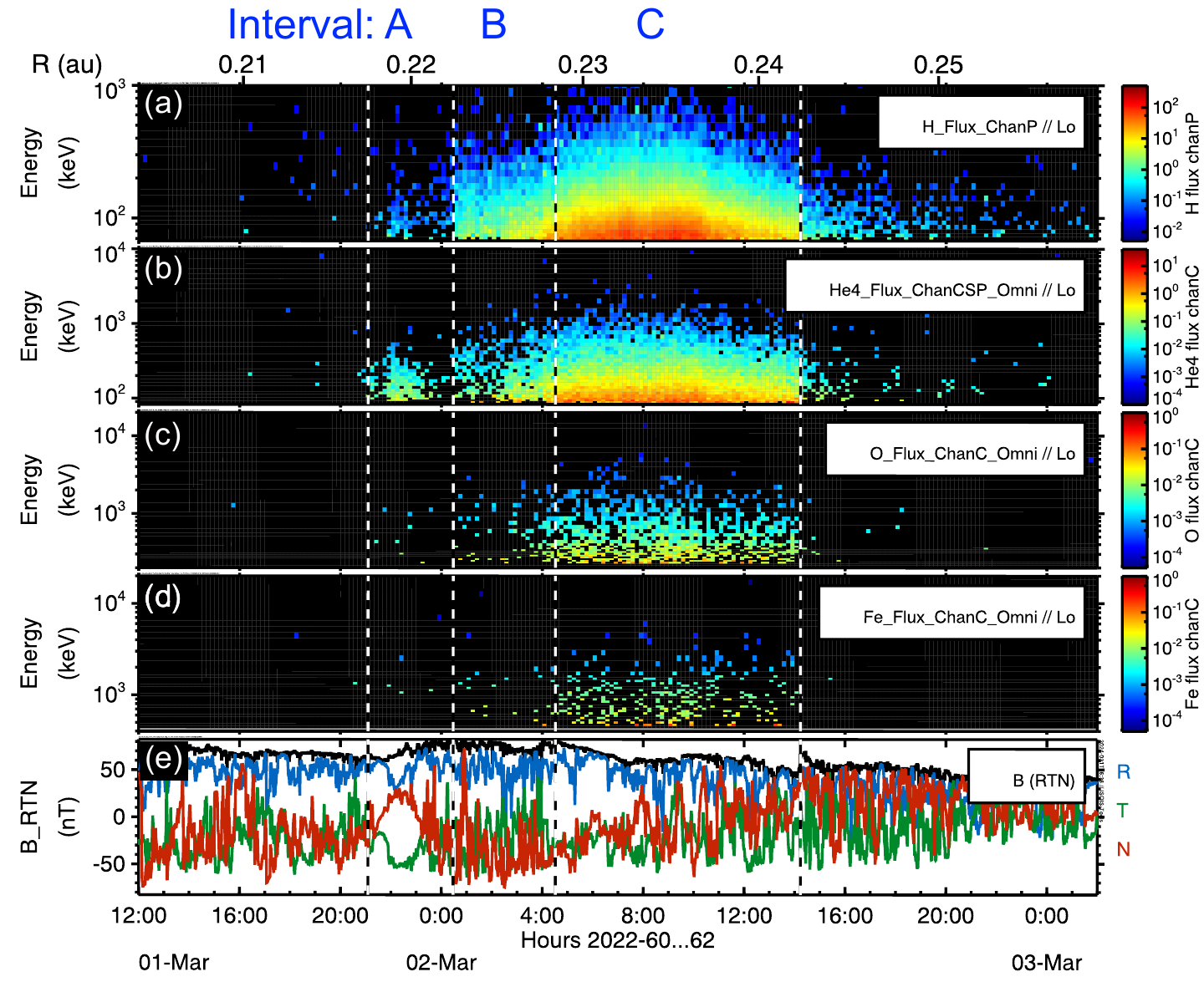}
  \caption{
Compositional signatures observed throughout the March 2, 2022 event. Enhancements are most pronounced in the CME  (Interval C), and we observe enhancements in H, He, O, and Fe  (a) - (d). Magnetic field vectors in panel (e) are shown for context   }
  \label{fig:f6}
  \end{figure}

The compositional signatures (Figure \ref{fig:f6})  are striking throughout the event. We observe elevated fluxes of H,$^4$He, O, and Fe. There are also indications of enhanced  $^3$He as well, particularly in the CME, but less than 1\% of $^4$He fluxes. These compositional variations are analyzed in greater detail  (see \S \ref{sec:spectra}) using the energy spectra.

The suprathermal heavy ions observed  begs the question of whether flaring  produces the majority of the suprathermal seed populations. A population from flares should be accompanied by strong anisotropies and speed dispersion due to particle propagation. However, these anisotropy dispersion signatures are largely absent in Interval C. 

\begin{figure}[ht]
    \centering
\includegraphics[width=\textwidth]{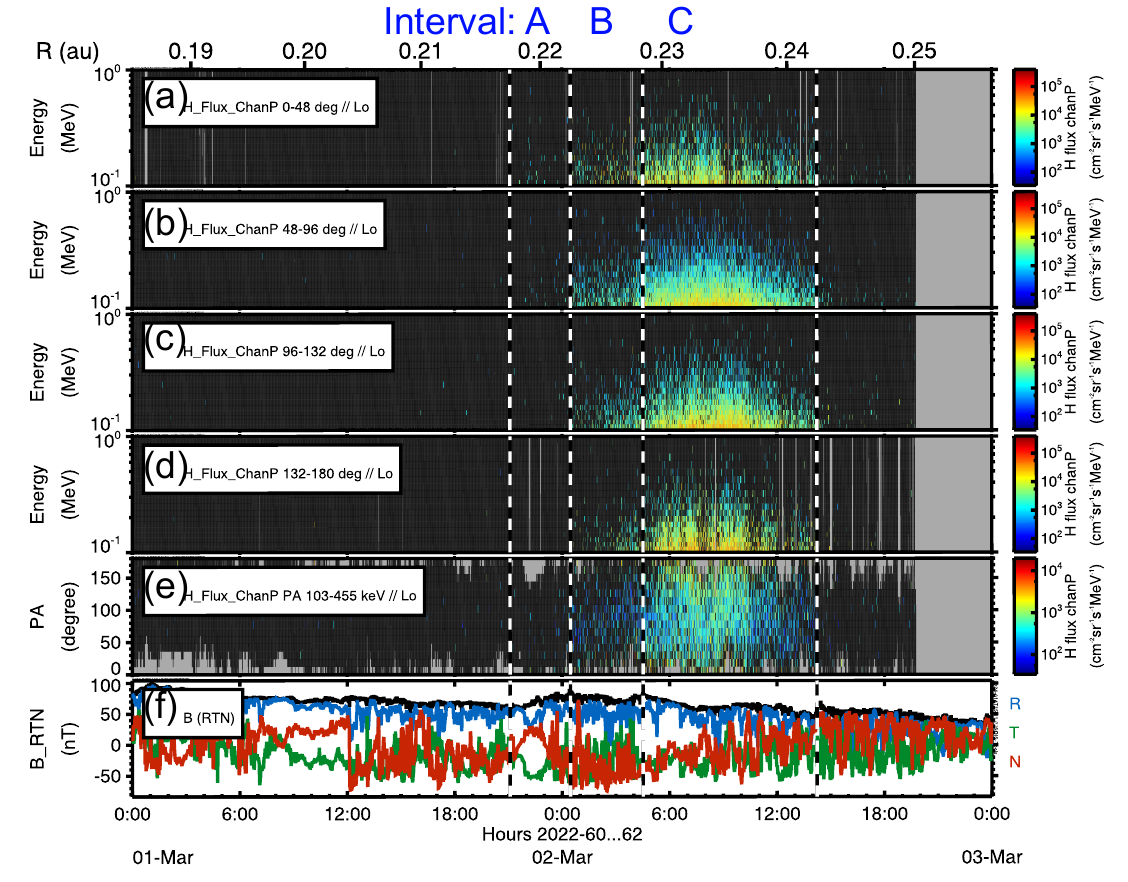}
  \caption{
Anisotropies observed in H$^+$ throughout the March 2, 2022 event. Energy spectra in pitch-angle bins $0^\circ -- 48^\circ$ (Panel a), $48^\circ -- 96^\circ$ (Panel b), $96^\circ -- 132^\circ$ (Panel c), and $132^\circ --180^\circ$ reveal strongly anisotropic distributions only in interval A (the flux rope). The pitch-angle distribution for ions with energy less than 200 keV shows that the turbulent sheath (interval B) and the CME (Interval C) have pitch-angle distributions almost symmetric about $90^\circ$.  The turbulent sheath distribution indicate localized acceleration, scattering  near $90^\circ$, and the near isotropic distribution internal to the CME  suggest trapping within the closed structure, and possibly acceleration mechanisms that act locally. Context from field measurements is shown in Panel D.    }
  \label{fig:f7}
  \end{figure}

\begin{figure}[ht]
    \centering
\includegraphics[width=\textwidth]{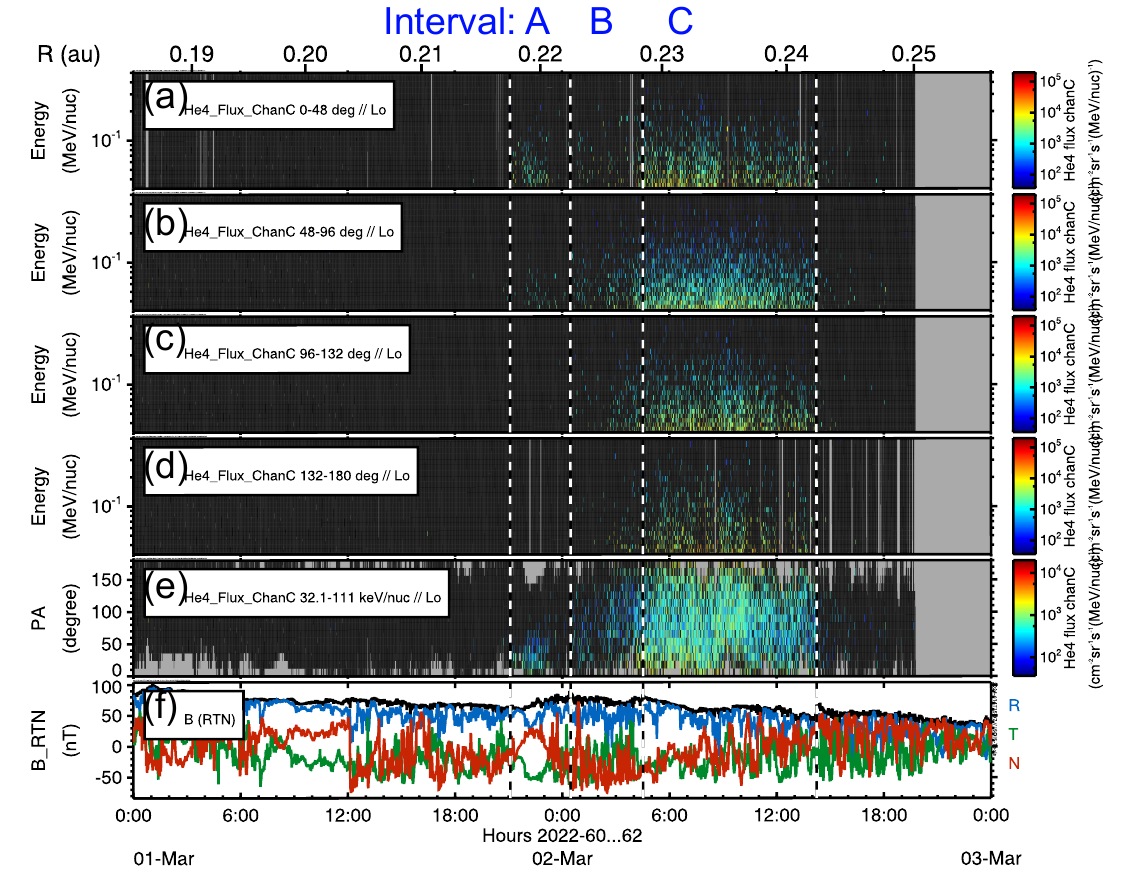}
  \caption{
Anisotropies observed in He ions throughout the March 2, 2022 event. The Figure format is identical to Figure \ref{fig:f7}, but shown here He ions instead of H$^+$. He ions consistently show strong anisotropies in the flux tube but distributions roughly centered on $90^\circ$ pitch-angle Intervals B and C.
}	  
  \label{fig:f8}
  \end{figure}

Figures \ref{fig:f7} and \ref{fig:f8} show the anisotropies observed for H$^+$ and He ions. Outward strong anisotropies are observed in Interval A, which is consistent with a particle source much closer to the Sun. However, Intervals B and C show pitch-angle distributions roughly symmetric about $90^\circ$. This is an indication of localized scattering, trapping and acceleration of the particle distributions. The presence of elevated fluxes  of heavy ions together with localized scattering  presents a unique signature  tied to trapping of particles within  the closed CME flux tubes \cite[]{Marsden:1987}.

\begin{figure}[ht]
    \centering
    \includegraphics[width=\textwidth]{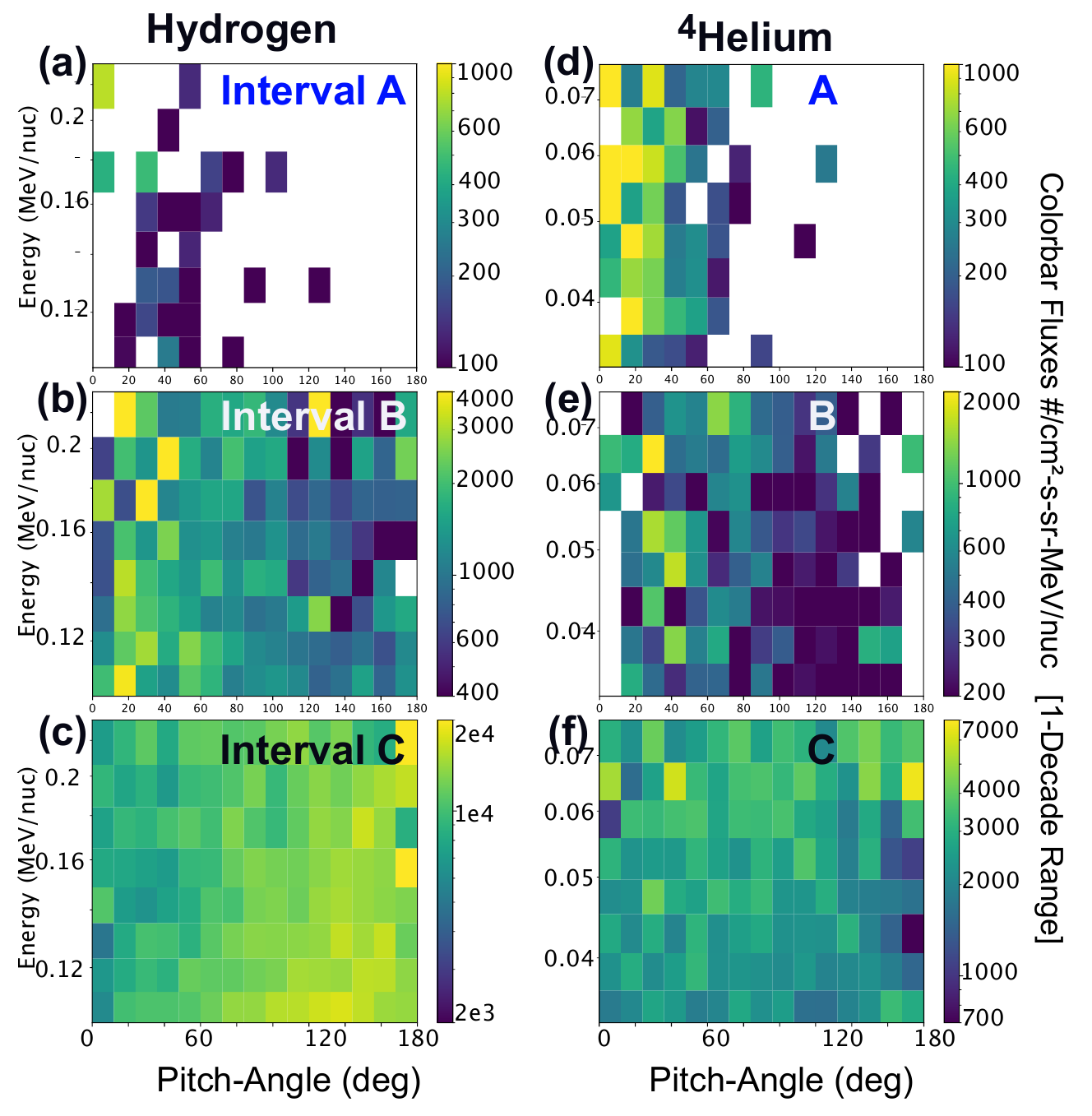}
  \caption{
Energy distributions as a function of pitch-angle of protons (Panels a, b, and c) and $^4$He (Panels d, e, f). Color coding in each panel is based on differential flux measurements.    
In order to emphasize the pitch-angle variability as a function of energy, the dominant spectral dependence of flux on energy has been removed: all fluxes are scaled proportional to the energy to the 3.2 power, maintaining the absolute flux values at 0.1 MeV/nuc.
}	  
  \label{fig:f9}
  \end{figure}

Energy versus pitch-angle spectragrams shown in Figure \ref{fig:f9} reveal where these particles are likely accelerated. In order to emphasize the pitch-angle variability as a function of energy, the dominant slope dependence of flux on energy has been removed: all fluxes are scaled proportional to the energy to the 3.2 power, maintaining the absolute flux values at 0.1 MeV/nuc.

Interval A (Panels a, and d) is associated with the isolated flux rope and shows pitch-angle distributions with particles directed predominantly outward from the Sun. This outward ion streaming indicates the acceleration of these ions must have occurred closer to the Sun.  

At the lowest energies ($\sim$ 100 keV for H and $\sim$ 40 keV/nuc for He) in Interval B (Panels b, and e), the sheath of the CME, the bulk of the population shows ion streaming away from the Sun.  At increasing energies, the populations become more isotropic, indicating inhibited transport due to scattering. However, even at the highest energies shown (240 keV for H   and 75 keV/nuc for He), the distribution remains somewhat anisotropic, with the majority of ions still coming from the Sun. 

Interval C in Figure \ref{fig:f9} is associated with the CME itself. 
For the protons, as energy decreases from 200 keV - 100 keV, we observe the distribution to evolve from an inward streaming distribution, with particles streaming \emph{back toward the Sun},  to a distribution that becomes increasingly isotropic. Note that the pitch-angle distributions are shown in the spacecraft reference frame, and are even larger in the solar wind reference frame.  Even at the lowest energies for protons ($\sim$ 100 keV), the distribution shows some inward streaming. 

While the first adiabatic invariant is not purely conserved due to scattering, the effect of ions streaming in toward the Sun drives the distribution progressively toward 90$^\circ$ pitch-angle. The effect becomes most pronounced at relatively low energies where the scattering time increases as particles are driven more strongly by higher field strengths closer to the Sun. 

For $^4$He,  at energies above 60 keV/nuc, the distribution is almost isotropic. Below 60 keV/nuc, the $^4$He shows an overall anisotropy with particles streaming \emph{outward } from the Sun. While the sense of the outward anisotropy is the same observed in Intervals A and B, the distribution is much closer to being isotropic in Interval C. 

Together, these observations suggest that ions are entrained within the fields of CME. Protons show inward streaming, and both proton and $^4$He distributions are much closer to being isotropic than in Interval A and B. The effects observed suggest that the magnetic field and flux tubes of the CME contain these particle distributions, acting as a reservoir for the ions that enter the closed field structures.

The inward streaming of protons suggests that the source for these protons must exist either further out from the Sun, or from the opposite flank of the CME.  Compression at the nose of the CME may accelerate protons that  can travel back toward the Sun to PSP at the flank of the CME. 

 The fluxes of protons, $^4$He, O, and Fe are all significantly enhanced throughout Interval C (Figure \ref{fig:f6}). Moreover, the average flux distributions all extend to the highest energies roughly 4 hours into Interval C where we observe large rotations in the magnetic field (Figure \ref{fig:f2}, panel c) and a bipolar structure is observed  in the helicity (Figure \ref{fig:hmpvi}, panel c). 
 
The closed field CME  acts as a magnetic bottle. While the energetic particles are effectively entrained in the fields at the flanks and front of the CME, these particles must also be significantly reflected at its legs to be  contained. In this context, the observations by \cite{McComas:2023} of the leg of a CME showing the complete absence of energetic and suprathermal ions is important, as it partially confirms that the legs of the CME with very strong fields must act to reflect ions. 

\section{Magnetic Field Turbulence and the Absent Signatures of  Rapid Wave-Particle Acceleration }

In this section, we show results of analysis of the magnetic turbulence characteristics and wave power. Increased wave power, particularly in Interval C, could indicate local particle acceleration. However, the observations show the \emph{absence} of such signatures, indicating that it is unlikely  that wave-particle acceleration is enhanced in the CME (Interval C). 

\begin{figure}[ht]
    \centering
    \includegraphics[width=\textwidth]{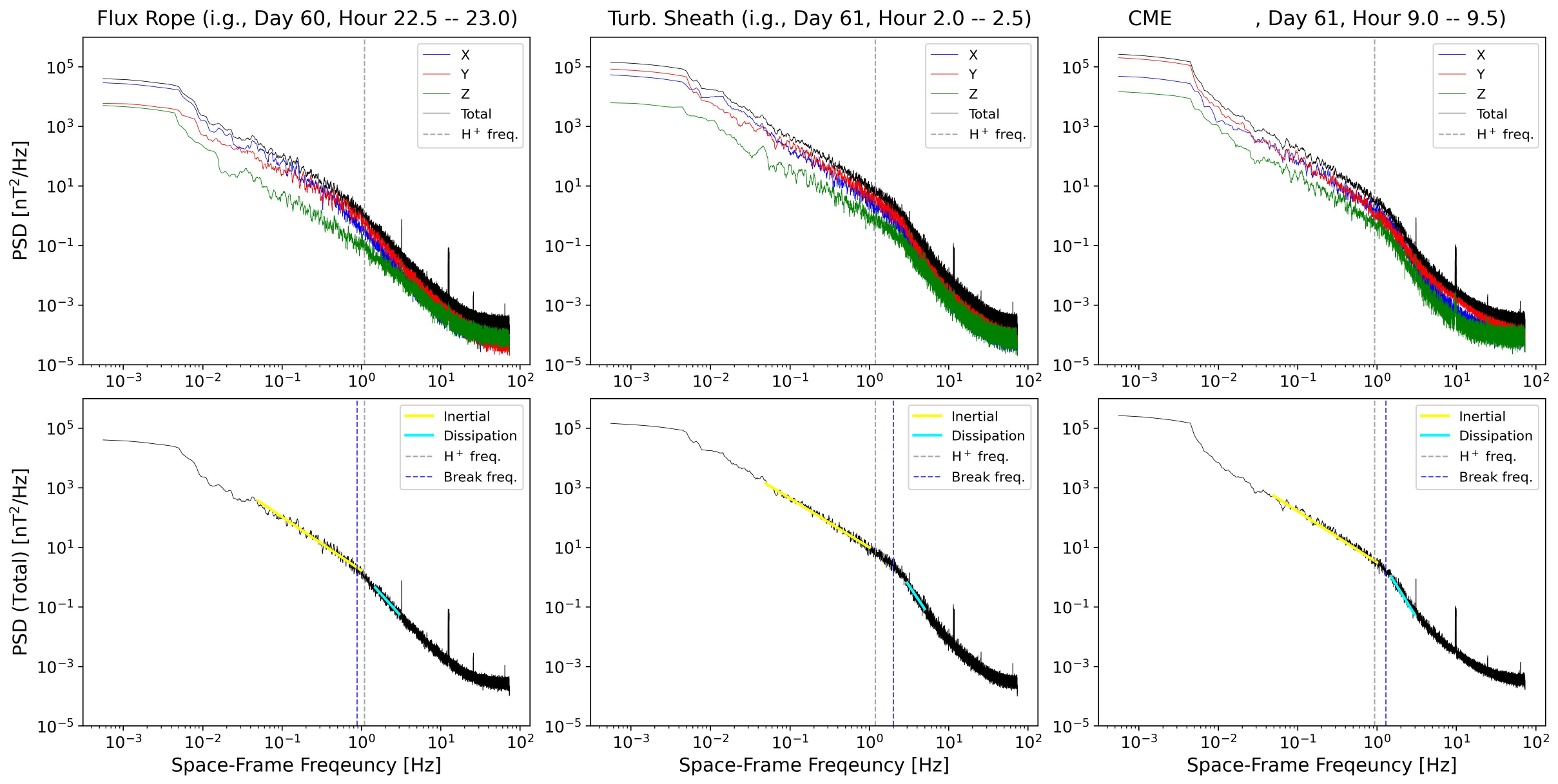}
  \caption{
Examples of  fits to the observed power spectrum $P = A f^n$ over the frequency intervals $0.01 - 1$ Hz for the inertial range and $1.5-3$ Hz for the dissipation range.  We also show the gyrofrequency of $H^+$ and the break in the spectrum where the inertial range and dissipation rang fits intersect. The top row of panels shows the $X$, $Y$, and $Z$ components of the power-spectral density (PSD), and the bottom row of panels shows the total PSD power in each frequency interval. The three columns of panels show the flux rope (interval A), the turbulent sheath (interval B) and the CME  (interval C). 
}	  
  \label{fig:f10}
  \end{figure}

Figure \ref{fig:f10} shows examples of power-spectral density (PSD) in the magnetic field and fits to these spectra:
\begin{itemize}
\item Power spectrum densities are formed over each 30-min interval;
\item Inertial fits are performed over the frequency range, from 0.01 Hz to 1 Hz, using $P = A\times f^n$ where $A$ is the amplitude parameter and $n$ is the power-law index;
\item Similarly, dissipation fits are performed over two different ranges:
\begin{itemize}
\item from 1.5 Hz to 3.0 Hz  in the  flux rope (Interval A) and within CME  (Interval C) intervals;
\item	from 3.0 Hz to 5.0 Hz in  turbulence sheath interval (Interval B).
\end{itemize}
\item	Proton gyration frequencies were calculated using the magnetic field data;
\item 	Break frequencies are found at which the inertial and dissipation range fits intersect.
\end{itemize}
The PSD spectra were computed using a Blackman-Tukey analysis that first computes autocorrelation matrix from the 3 measured components of the magnetic field 
\cite[]{blackman1958measurementI, matthaeus1982measurement}.  The power spectrum is the Fourier Transform of the autocorrelation matrix with the total power being the trace of the spectrum for the 3 components.

The low frequency interval $\sim 0.01$ Hz is  suitable to describe the power spectral density at large inertial scales (as opposed to small dissipation scales) since its value is significantly smaller than the proton cyclotron frequency (typically of order 1 Hz). The high frequency interval $\sim 3$ Hz, on the other hand, is chosen to describe smaller scale dissipative motions.

In both cases, Interval B shows a sudden increase of the power of magnetic fluctuations relative to the adjoining flux rope and CME, consistent with observations (Figure 2) of increased solar wind speed, temperature, and field strength in the  turbulence structures (Interval B). Furthermore, the  steepening of the dissipation range spectra in this interval (the bottom figure) is consistent with elevated turbulence levels \cite[]{Smith:2006}, indicating that the charged particles inside the plasma become more diffusive and experience more rapid scattering. The  increase in scattering in regions of a speed gradient could provide for more rapid energization, as typically found in cases of strong diffusive acceleration at a shock or a compressive structure.

\begin{figure}[ht]
    \centering
    \includegraphics[width=\textwidth]{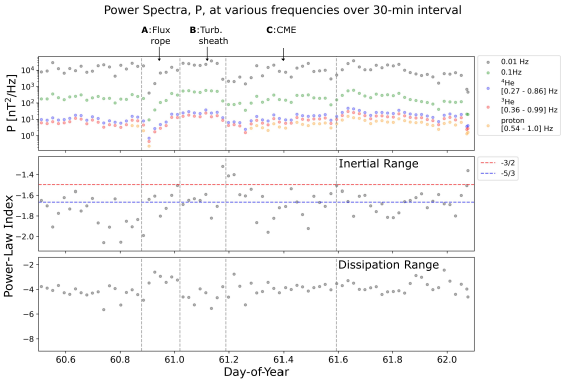}
  \caption{
The power per Hz in the PSD spectra (top panel) and   results of the power-law fits (middle and bottom panels in each 30-min interval). 
 The top panel shows the power spectral densities at 0.01 Hz and 0.1 Hz, respectively, together with power in gyrofrequency intervals associated with $^4$He, $^3$He, and protons. Each fit was performed over the frequency range from 0.01 Hz to 1 Hz, using $P = A f^n$, where $A$ is the amplitude parameter and $n$ is the power-law index.}
  \label{fig:f11}
  \end{figure}

Near the beginning of Interval A (the flux rope) and Interval C (CME) in Figure \ref{fig:f11}, we observed a pronounced decrease in turbulence power (this is most evidence in the inertial range). In Interval C, the drop in interval range power also corresponds with a flattening of the power-law index. Both signatures are consistent with the effects of expansion seen in Figure \ref{fig:f2}: in the first 3 hours of Interval C, we see a rapid reduction in the field strength, a reduction in the particle speed and temperature.

Similarly, in the first several hours of Interval A (the flux rope) we see the plasma density decrease, and a drop in the plasma temperature, which suggest plasma expansion. Therefore, the effects of plasma expansion near the beginning of Interval A and Interval C could explain the reductions in turbulence power. 

The turbulence power after the first few hours in Interval C appears to recover to nominal values  over this period. We do not observe higher levels of turbulence that might explain local particle acceleration. While this appears to be a negative conclusion, it is important to highlight the lack of signatures in EMIC waves that would be expected if the physics conforms to an auroral pressure cooker mechanism. It is precisely the \emph{absence} of these signatures that point to other sources of particle trapping discussed further in \S \ref{sec:discuss}. 

The confinement of protons, $^3$He, $^4$He and heavy ions is observed in the flank of the CME: ions accelerated at $<1$ MeV/nuc have pitch-angle distributions that do not indicate strong streaming from the Sun or back toward the Sun (see Figure \ref{fig:f9}, panel c). 

\section{Energetic Particle Spectra: Trapping, Cooling in the CME, and Compressive or Shock Acceleration beyond}
\label{sec:spectra}

The distinction observed in anisotropies between the flux rope (Interval A), the turbulent sheath (Interval B) and the CME (Interval C) indicates that scattering or trapping is dominant in the CME, whereas the flux rope and turbulent sheath are largely sourced from acceleration occurring closer to the Sun or from flare populations. We provide an examination of the particle energy spectra to gain further insight into these differences.

\begin{figure}[ht]
    \centering
    \includegraphics[width=\textwidth]{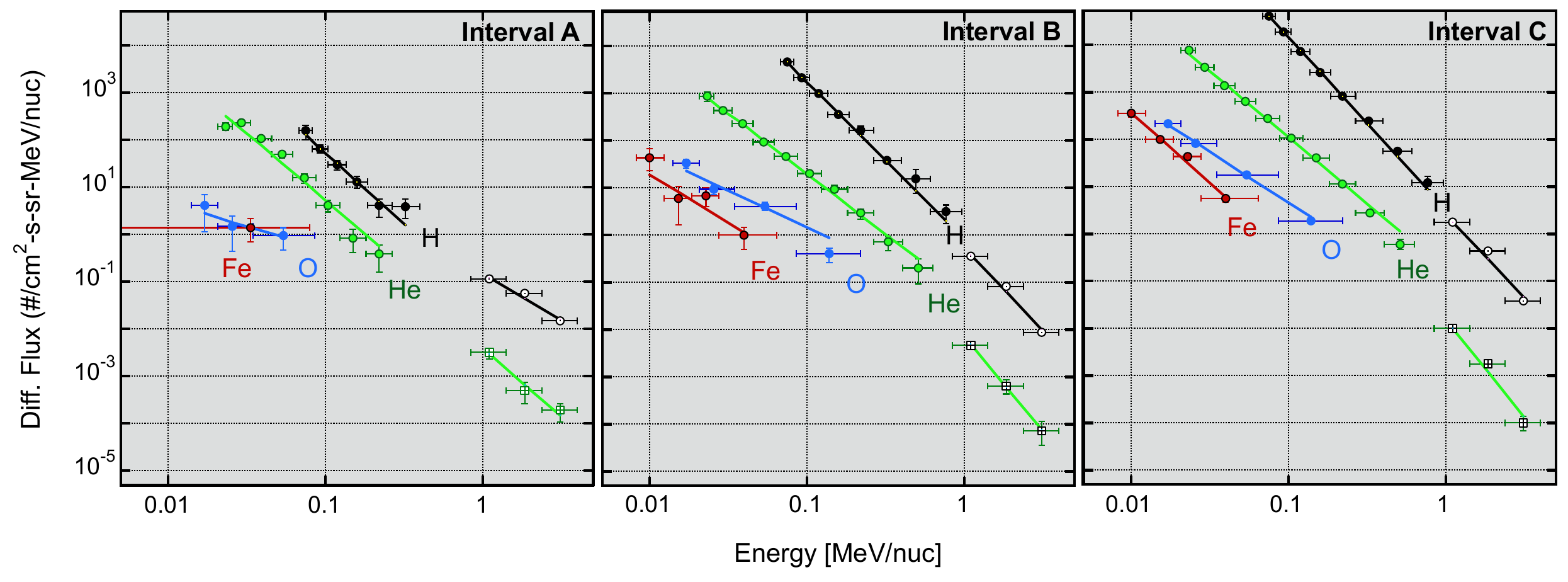}
  \caption{
Energetic particle spectra and spectral slopes in each interval. Spectral slopes are shown above and below 1 MeV/nuc for H, He, O, and Fe. Fe and O fits are not shown above 1 MeV/nuc due to poor statistics. 
}
  \label{fig:f12}
  \end{figure}
  
  \begin{figure}[ht]
    \centering
    \includegraphics[width=\textwidth]{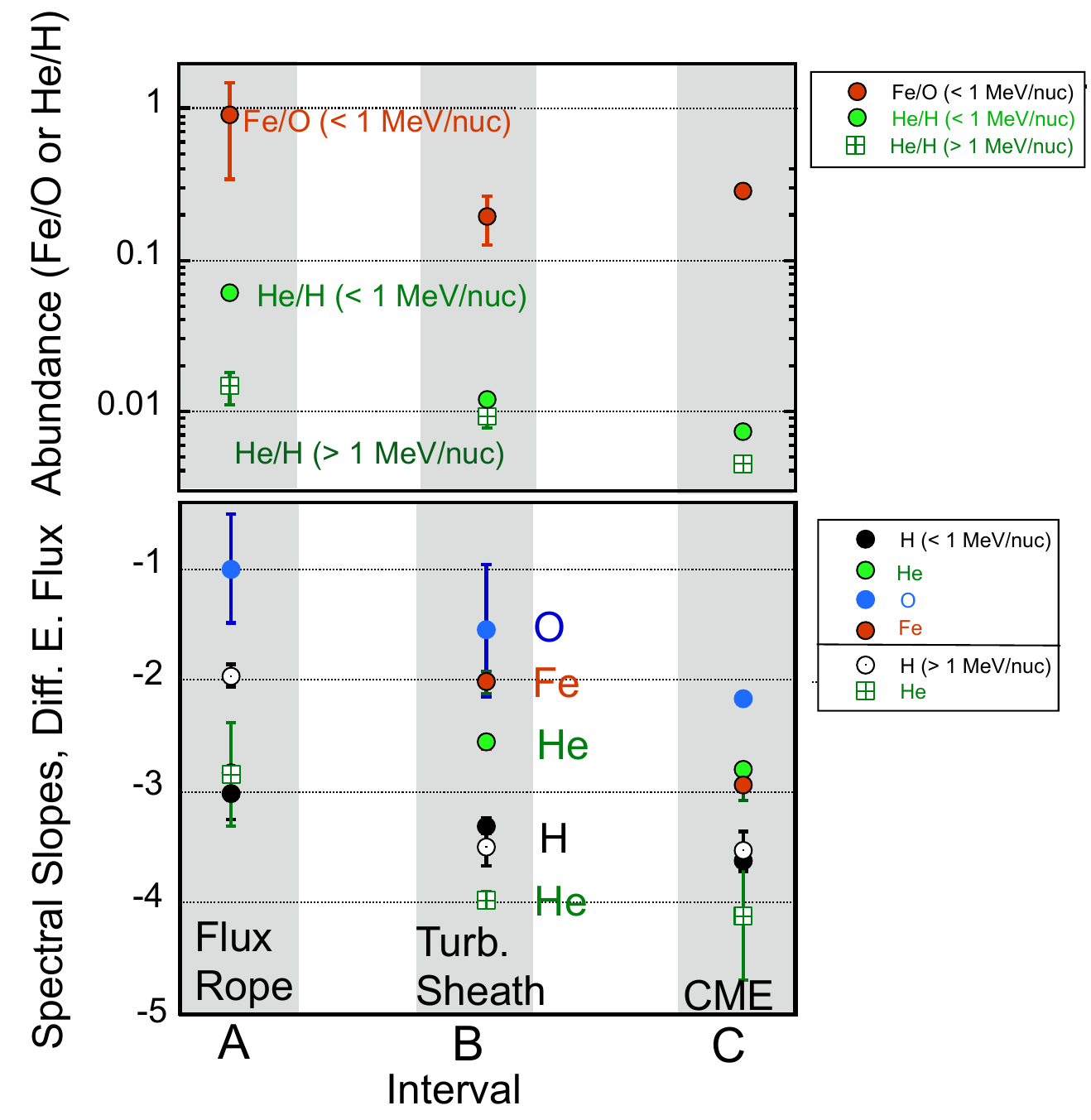}
  \caption{
Comparison of spectral slopes for differential energy flux and abundance ratios as a function of the observed interval. 
}
  \label{fig:f13}
  \end{figure}

Figure \ref{fig:f12} shows the spectral slopes derived from H (black), He (green), O (blue) and Fe (rust) and Figure \ref{fig:f13} compiles these power-laws and abundance ratios (Fe/O and He/H) within the intervals observed. Table \ref{tab:Abundance} also lists the Fe/O and He/H abundance ratios for the intervals. 
Figure \ref{fig:f13} shows a clear progression of power-laws from the three intervals from hardest spectra in the flux rope (Interval A) to the softest spectra in the CME  (Interval C). This progression mimics that observed in anisotropies: the largest anti-sunward anisotropies in Interval A are associated with the hardest spectra; whereas the weakest anisotropies (or even slightly sunward streaming) are associated with the softest power-law distributions. As distributions are scattered in the CME sheath and contained within the CME reservoir, the dwell times increase within the plasma. Adiabatic cooling acts preferably on ions with significant energy components perpendicular to the field. In contrast, ions streaming predominantly along the field have low energies perpendicular to the field and are transported without necessarily undergoing large amounts of cooling.  While we see the highest total fluxes in the CME, these particles have experienced longer dwell times as they are contained within the closed CME structure. The cooling of these distributions occurs as the CME expands and the fields weaken, causing the spectral slopes to soften significantly. 

The energy spectra have been plotted with differential flux as a function of energy per nucleon. This is appropriate for mass-independent acceleration mechanisms. However, acceleration through motion across electric fields should order the differential fluxes differently, as a function of energy-per-charge. There are also  balances that develop between particle partial pressures including the suprathermal particle pressure and magnetic field pressure in the system, which would suggest an ordering of the data as a function of net energy. Results of  distributions ordered in these ways are being actively studied, particularly in how they inform particle acceleration and the roles of flux-tubes and transport effects in channeling particle distributions.

The abundance ratios show a similar pattern as observed in the spectral slopes. Interval A stands out with much higher He/H and Fe/O ratios below 1 MeV/nuc, which clearly indicates a flaring source for the suprathermal population below 1 MeV/nuc. However, above 1 MeV/nuc the He/H abundance ratio in Interval A suggests a combination of sources from flares and accelerated plasma. 

Interval B and C show a progression of lower He/H abundances suggesting that the sheath and CME contain increasing amounts of accelerated plasma. Similarly, the Fe/O abundance ratio is lower in Interval B and C compared to Interval A. 

Taken together, the spectral slopes, the anisotropies and the abundances are consistent: Interval A appears indicative of a flare suprathermal source with large outward anisotropies, harder spectral slopes, and large He/H and Fe/O abundance ratios below 1 MeV/nuc; Interval B and C indicate increasing amounts of scattering and acceleration from the plasma which leads to progressively more isotropic distributions,  softer spectral slopes, and decreasing abundance ratios.  

\begin{deluxetable}{cccc}[ht]
\tablecaption{Abundance Ratios (Fe/O and He/H in the March 2, 2022 event \label{tab:Abundance}}
\tablehead{
\colhead{Interval} & \colhead{Fe/O} &  \colhead{He/H} &  \colhead{He/H}  \\
\colhead{} & \colhead{0.014-0.08 MeV/n} & \colhead{0.07-0.75 MeV/n} & 
\colhead{1-4 MeV/n} 
 }
\startdata
A     		& $0.908\pm 0.568$  & $0.061\pm 0.015$ &  $0.0146 \pm 0.0035 $  \\
B     		& $0.194\pm 0.067$  & $0.012\pm 0.001$ &  $0.0093 \pm 0.0016 $  \\
C     		& $0.289\pm 0.026$  & $0.0073\pm 0.0002$ &  $0.0045 \pm 0.0004 $  \\
 \hline
 \enddata
\end{deluxetable}

From the power-laws in Figure \ref{fig:f13}, we infer another important signature.  For energies $> 1$ MeV/nuc in both the turbulent sheath and the CME,  the H and He spectra converge to the same power-law. This is precisely what is expected in diffusive shock acceleration (DSA), and suggests that the CME   is connected to a weak shock or a strong compression further from the Sun. 




\section{Discussion of Particle Sources and Acceleration in the Isolated Flux Rope, the Sheath and CME}
\label{sec:discuss}

We have examined  a rare event on the  flank of the CME where we observe the passage of a flux tube, a turbulent sheath, and the  CME. In the following subsections we summarize the observations in each of the intervals and structures studied, and tie these to source  mechanisms acting on the particle distributions. 

\subsection{Interval A: the isolated flux rope}

\begin{deluxetable}{ccccccc}[ht]
\tablecaption{Truth table for acceleration mechanisms from \ISOIS~ observations within Interval A, the isolated flux tube. \label{tab:A}} 
\tablehead{
\colhead{Obs.} & \colhead{Fig.} &  \colhead{Energy}  & \colhead{Spec} & \colhead{Trend} & \colhead{Exp/} & \colhead{Source}   \\
\colhead{} & \colhead{} & \colhead{MeV/n} & \colhead{} & \colhead{} & \colhead{Comp} & \colhead{} 
 }
\startdata
$|B|$     		& 5d   &        						&           &    $\nearrow$ &   Comp    &         \\
$V$     			& 5g   &           & H$^+$            &    $\nearrow$ &   Comp     &          \\
Dens     		& 5e   &          & H$^+$            &     $\uparrow$$\downarrow$     &   &       \\
Temp.     		& 5f    &         &  H$^+$            &    $\nearrow$ &   Comp     &         \\
Bi-Dir    		& 6a   &           & e-            		 &    $\uparrow$ &         & Closed B     \\
B Rot\tablenotemark{a}        & 8a   &      &       &   $\uparrow$ (1-rot) &      &      Corona     \\
H$^+$    	& 5a    &  0.1-0.5         &    H$^+$        &    $\downarrow$ &       &       \\
He/H    	& 18    &  0.07-0.75         &       He,H   &    $\uparrow$ &       &   Flares    \\
He/H    	& 18    &  1-4         &       He,H   &    $\uparrow$ &        &   Flares, Plasma    \\
Fe/O & 18   &  0.014-0.08         & Fe,O  &    $\uparrow$ &     &        Flares \\
Anisotropy     & 14d &  0.02-0.07  &   $^4$He   & antisun         &     &       $<$0.2 au    \\
Spect Ind\tablenotemark{b}       &   18 &   $<$1   & H,He,O,Fe &    $\uparrow$Distinct &       &       \\
Spect Ind\tablenotemark{c}   & 18  & $>$1  &  H,He   &    $\uparrow$Distinct &         &      \\
Turb Pwr   & 16a  &      &          &    $\nearrow$ & Comp     &         \\
PVI    & 8b   &       &          &    $\nearrow$ &      &    \\
Helicity Pk\tablenotemark{d}          & 8d   &      &        &    $\rightarrow$ &       &          \\
Suprathermal-EPs\tablenotemark{e}          & 5b,c   & keV-MeV      &  H      &    $\downarrow$ &       &          \\
\hline \\
\colhead{Mechanism} & \colhead{Score} & \colhead{MeV/n} & \colhead{Assessment} & \colhead{Source} & \colhead{} & \colhead{} 
 \\
 \hline
DSA local\tablenotemark{f} & 1/2		& $<1$ MeV/n  & No   \\  
DSA nonlocal\tablenotemark{g} & 0/1  & $>$ 1 MeV/n & No & \\
\textbf{Flare Src}\tablenotemark{h}			& 			\textbf{4/4}	 & $<1$ \textbf{MeV/n}	&  \textbf{Yes}	&  Active rgn, \\
		& 			 & 	&  	& Reconfig B \\
Trapping/Confinement\tablenotemark{i}		& 		0/3  			 & $<1$ MeV/n	&  No \\
Acc nonlocal\tablenotemark{j}	& 		2/4  			 & $<1$ MeV/n	&  No & \\
\enddata
\tablenotetext{a}{Rotation(s) in the magnetic field occupying a large portion of the interval. We observe 1 large rotation in the field}
\tablenotetext{b}{Spectral slope quite large -3 to -1.5 and distinct for species}
\tablenotetext{c}{Spectral slope large -3.5 to -2 and distinct for species}
\tablenotetext{d}{Significant helicity peak for $\ell/\lambda = 3$. }
\tablenotetext{e} {Connected spectrum from keV suprathermal to MeV energetic protons } 
\tablenotetext{f} {2 tests for local Diff. Shock Accel (DSA) at shock or compression: velocity gradient, common spectrum for H, He, O, Fe} 
\tablenotetext{g} {One test for non-local DSA: common spectral slope for H, He} 
\tablenotetext{h} {Four tests for flare source: antisunward anisotropy, hard spectra, high He/H ($>0.004$), high Fe/O ($> 0.124$)   } 
\tablenotetext{i} {Three tests for trapping/confinement: enhancement from suprathermal to EPs, weak anisotropy, lower spectral indices from cooling} 
\tablenotetext{j} {Four tests for non-local acceleration, not DSA: no common spectrum for species, strong anisotropy, low He/H, low Fe/O} 
\end{deluxetable}

Table \ref{tab:A} provides a summary of observations and a truth table for source mechanisms. For Interval A, the presence of bi-directional electrons, and a large rotation in the magnetic field indicates a  closed flux rope. The strong anti-sunward anisotropy indicates a source for lower energy ($< 1$ MeV/nuc) energetic ions inside 0.2 au. 

The distinct spectral indices observed for H, and He show that the acceleration mechanism does not create single spectral form. This is a strong indicator that DSA is not a viable source mechanism. The  relatively high Fe/O ratio and high He/O ration, particularly below 1 MeV/nuc,   makes it likely that flares  are the sources for the energetic ions. A flare source for the energetic ions below 1 MeV/nuc is also consistent with the high spectral indices (hard spectra) observed (between -1 and -3 for O, H and He) in this interval.  


 Interval A likely shows suprathermal ions produced from flaring associated with recent reconfiguration of the flux tube. Further, rapid expansion of the flux tube could effectively cool populations that are not flare-like, and the flare-produced suprathermal populations will come to the observer more quickly, even without significant dispersion. This suggests, as indicated in the first appendix that the flux tube is undergoing  rapid time-dependent changes. 

\subsection{Interval B: Turbulent Sheath}

The turbulent sheath shows enhanced levels of turbulence, and the absence of bi-directional electrons implies that the magnetic structure is open and simply connected (Table \ref{tab:B}). The signatures of compression in the plasma temperature, magnetic field and turbulence power all indicate that the sheath is compressed solar wind plasma leading the CME. 

We observe  levels of Fe/O levels that, while somewhat larger  than 0.134 (the Fe/O typical of SEPs), are not significantly higher.  The He/H ratios are lower than in Interval A, but still higher than values typical of SEPs ($\sim 0.004$).  Further, we observe a large antisunward anistropy. These indicators suggest that both flaring and accelerated plasma sources contribute to the energetic particle populations.  

The almost common spectral slope for H, He at higher energies ($>$ 1 MeV/n) provide a signature of DSA. The fact that this similar  spectral slope occurs at higher energies but remains distinct below 1 MeV/n indicates that the DSA process is nonlocal, and beyond the point of observation.  Note that the spectral slope $-\gamma$ is related to the shock compression, 
\begin{eqnarray}
\gamma = \frac{r_c + 2}{2(r_c - 1)} .
\end{eqnarray}
The spectral slope of $\gamma \sim 4$ indicates a weak shock, $r_c \sim 1.4$, which is  consistent with the slow moving CME and therefore weak compression. 

The results derived from the truth table (Table \ref{tab:B}) suggest the classic case of a low energy flare and accelerated plasma seed population that is fed into DSA at higher energies. 

\begin{deluxetable}{ccccccc}[ht]
\tablecaption{Truth table for acceleration mechanisms from \ISOIS~ observations within Interval B, the turbulent sheath. \label{tab:B}} 
\tablehead{
\colhead{Obs.} & \colhead{Fig.} &  \colhead{Energy}  & \colhead{Spec} & \colhead{Trend} & \colhead{Exp/} & \colhead{Source}   \\
\colhead{} & \colhead{} & \colhead{MeV/n} & \colhead{} & \colhead{} & \colhead{Comp} & \colhead{} 
 }
\startdata
$|B|$     		& 5d   &        						&           &    $\uparrow$ &   Comp    &         \\
$V$     			& 5g   &           & H$^+$            &    $\uparrow$ &       &          \\
Dens     		& 5e   &          & H$^+$            &     $\uparrow$     &   &       \\
Temp.     		& 5f    &         &  H$^+$            &    $\uparrow$ &   Comp     &         \\
Bi-Dir    		& 6a   &           & e-            		 &    $\downarrow$ &         & Open B     \\
B Rot\tablenotemark{a}        & 8a   &      &       &   $\downarrow$  &      &      Sol Wind     \\
H$^+$    	& 5a    &  0.1-0.5         &    H$^+$        &    $\nearrow$ &       &       \\
He/H    	& 18    &  0.07-0.75         &       He,H   &    $\nearrow$ &       &   Flares, Plasma    \\
He/H    	& 18    &  1-4         &       He,H   &    $\nearrow$ &        &   Flares, Plasma    \\
Fe/O & 18   &  0.014-0.08         & Fe,O  &    $\rightarrow$ &     &       Flares, Plasma \\
Anisotropy     & 14b &  0.1-0.2  &   H   & antisun         &     &       $<$0.2 au    \\
Anisotropy     & 14e &  0.03-0.07  &   $^4$He   & antisun         &     &       $<$0.2 au    \\
Spect Ind\tablenotemark{b}       &   18 &   $<$1   & H,He,O,Fe &    $\uparrow$Distinct &       &       \\
Spect Ind\tablenotemark{c}   & 18  & $>$1  &  H,He   &    $\searrow$Same &         &  DSA, $> $0.2 au    \\
Turb Pwr   & 16a  &      &          &    $\uparrow$ & Comp     &         \\
PVI    & 8b   &       &          &    $\uparrow$ &      &    \\
Helicity Pk\tablenotemark{d}          & 8d   &      &        &    $\rightarrow$ &       &          \\
Suprathermal-EPs\tablenotemark{e}          & 5b,c   & keV-MeV      &  H      &    $\downarrow$ &       &          \\
\hline \\
\colhead{Mechanism} & \colhead{Score} & \colhead{MeV/n} & \colhead{Assessment} & \colhead{Source} & \colhead{} & \colhead{} 
 \\
 \hline
DSA local\tablenotemark{f} & 0/2		& $<1$ MeV/n  & No   \\  
\textbf{DSA nonlocal}\tablenotemark{g} & \textbf{1/1} & $>$ 1 \textbf{MeV/n} & \textbf{Yes} & $>$0.2 au \\
Flare Src\tablenotemark{h}			& 			2/4	 & $<1$ MeV/n	&  No	& \\
Trapping/Confinement\tablenotemark{i}		& 		1/3  			 & $<1$ MeV/n	&  No \\
\textbf{Acc nonlocal}\tablenotemark{j}	& 		\textbf{4/4}  			 & $<1$ \textbf{MeV/n}	& \textbf{Yes} & $< 0.2$ \textbf{au} \\
\enddata
\tablenotetext{a}{Rotation(s) in the magnetic field occupying a large portion of the interval. We observe 1 large rotation in the field}
\tablenotetext{b}{Spectral slope quite large -3 to -1.5 and distinct for species}
\tablenotetext{c}{Spectral slope lower $\sim -4$ and the same for H and He}
\tablenotetext{d}{Significant helicity peak for $\ell/\lambda = 1$. No significant peaks found in Interval B}
\tablenotetext{e} {Connected spectrum from keV suprathermal to MeV energetic protons } 
\tablenotetext{f} {2 tests for local Diff. Shock Accel (DSA) at shock or compression: velocity gradient, common spectrum for H, He, O, Fe} 
\tablenotetext{g} {One test for non-local DSA: similar spectral slope for H, He} 
\tablenotetext{h} {Four tests for flare source: antisunward anisotropy, hard spectra, high He/H ($>0.004$), high Fe/O ($> 0.124$)    } 
\tablenotetext{i} {Three tests for trapping/confinement: enhancement from suprathermal to EPs, weak anisotropy, lower spectral indices from cooling} 
\tablenotetext{j} {Four tests for non-local acceleration, not DSA: no common spectrum for species, anisotropy, low He/H, low Fe/O} 
\end{deluxetable}

\subsection{Interval C: CME Flank}

The CME  shows lower levels of turbulence compared to Interval B, but these levels fluctuate and then increase mid-way through the Interval C  (Table \ref{tab:C}).  The bi-directionality of  electrons is extremely strong, showing a  closed magnetic flux tube. The signatures of expansion in the plasma temperature, and magnetic field  indicate that the CME is expanding as it moves through the region near 0.2 au. 

Throughout the flank of the CME we observe enhanced levels of $^3$He and of heavy ions in the energetic particles with a weak anistropy. The enhancements in energetic particle fluxes  connect all the way down in energy to enhanced suprathermal fluxes at $\sim$ keV energies. The spectral slopes are also steeper (softer spectra) within Interval C, as compared to Interval B, indicating softening of the suprathermal and energetic particle spectra as the CME expands and cools the suprathermal and energetic particle populations.  Together, these observations provide consistent indications that the energetic particles and suprathermal particles behave as a connected population, confined magnetically to the closed configuration of the flux rope CME. 

The anisotropies, though weak are mixed in direction with the protons generally showing a weak sunward streaming, and the $^4$He ions showing small indications of an anti-sunward streaming, particularly at energies between 32 keV/nuc and 60 keV/nuc. These signatures suggest both a flaring source near the Sun, and that there is additional acceleration out further from the Sun. The composition showing Fe/O and He/H enhancements but levels more typical of SEPs provide additional evidence of a mix of flare and plasma sources at lower energies ($<$ 1 MeV/n)  for the particle populations.     

As with Interval B, the similar spectral slope for H, He at higher energies ($>$ 1 MeV/n) likely shows that DSA is active. The similar spectral slope occurs at higher energies but remains distinct below 1 MeV/n indicating that the DSA process is beyond the point of observation$ >0.2$ au. The spectral slope of $-\gamma \sim -4$ also indicates a weak shock or compression, $r_c \sim 1.4$, and is consistent with compression in front of the slow moving CME. Note that the compression ratio needed is quite similar to  that found from Interval B. 

The results derived from the truth table (Table \ref{tab:C}) suggest the case of a low energy flare seed population that is fed into and contained by the CME. Therefore, the CME acts as a reservoir for these suprathermal and energetic particles. As in the case of the sheath (Interval B), there is likely DSA at higher energy occuring at a compression or shock near the front of the CME. 

\begin{deluxetable}{ccccccc}[ht]
\tablecaption{Truth table for acceleration mechanisms from \ISOIS~ observations within Interval C, the CME.  \label{tab:C}}
\tablehead{
\colhead{Obs.} & \colhead{Fig.} &  \colhead{Energy}  & \colhead{Spec} & \colhead{Trend} & \colhead{Exp/} & \colhead{Source}   \\
\colhead{} & \colhead{} & \colhead{MeV/n} & \colhead{} & \colhead{} & \colhead{Comp} & \colhead{} 
 }
\startdata
$|B|$     		& 5d   &        						&           &    $\searrow$ &   Exp    &         \\
$V$     			& 5g   &           & H$^+$            &    $\rightarrow$ &       &          \\
Dens     		& 5e   &          & H$^+$            &     $\uparrow$     &   &       \\
Temp.     		& 5f    &         &  H$^+$            &    $\downarrow$ &   Exp     &         \\
Bi-Dir    		& 6a   &           & e-            		 &    $\uparrow$ &         & Closed B     \\
B Rot\tablenotemark{a}        & 8a   &      &       &   $\uparrow$($>5$-rot)  &      &      Corona     \\
H$^+$    	& 5a    &  0.1-0.5         &    H$^+$        &    $\uparrow$ &       &       \\
He/H    	& 18    &  0.07-0.75         &       He,H   &    $\nearrow$ &       &   Flares, Plasma    \\
He/H    	& 18    &  1-4         &       He,H   &    $\rightarrow$ &        &   Flares, Plasma    \\
Fe/O & 18   &  0.014-0.08         & Fe,O  &    $\nearrow$ &     &        Flares, Plasma \\
Anisotropy     & 14c &  0.1-0.2  &   H   & $\sim$iso/snwrd         &     &       confined    \\
Anisotropy     & 14f &  0.03-0.07  &   $^4$He   & $\sim$iso/antisun         &     &       confined    \\
Spect Ind\tablenotemark{b}       &   18 &   $<$1   & H,He,O,Fe &    $\searrow\searrow$Distinct &       &       \\
Spect Ind\tablenotemark{c}   & 18  & $>$1  &  H,He   &    $\searrow\searrow$Same &         &  DSA, $> $0.2 au    \\
Turb Pwr   & 16a  &      &          &    $\nearrow$ &      &         \\
PVI    & 8b   &       &          &    $\nearrow$ &      &    \\
Helicity Pk\tablenotemark{d}          & 8d   &      &        &    $\uparrow$ &       &          \\
Suprathermal-EPs\tablenotemark{e}          & 10   & keV-MeV      &  H      &    $\uparrow$ &       &      CME plasma    \\
\hline \\
\colhead{Mechanism} & \colhead{Score} & \colhead{MeV/n} & \colhead{Assessment} & \colhead{Source} & \colhead{} & \colhead{} 
 \\
 \hline
DSA local\tablenotemark{f} & 0/2		& $<1$ MeV/n  & No   \\  
\textbf{DSA nonlocal\tablenotemark{g}} & \textbf{1/1}  & $>$ \textbf{1 MeV/n} & \textbf{Yes} &  $>0.2$ au\\
Flare Src\tablenotemark{h}			& 			2/4	 & $<1$ MeV/n	&  No	& \\
\textbf{Trapping/Confinement\tablenotemark{i}	}	& 		\textbf{3/3}  			 & $<1$ \textbf{MeV/n}	&  \textbf{Yes} \\
\textbf{Acc nonlocal}\tablenotemark{j}	& 		\textbf{4/4}  			 & $<1$ \textbf{MeV/n}	& \textbf{Yes} & $< 0.2$ \textbf{au} \\
\enddata
\tablenotetext{a}{Rotation(s) in the magnetic field occupying a large portion of the interval. We observe 1 large rotation in the field}
\tablenotetext{b}{Spectral slope quite large -3 to -1.5 and distinct for species}
\tablenotetext{c}{Spectral slope large -3.5 to -2 and distinct for species}
\tablenotetext{d}{Helical
structures at the scale of  $\ell/\lambda = 7$  appear within the CME, slight bipolar feature for $\ell/\lambda = 1$ }
\tablenotetext{e} {Connected spectrum from keV suprathermal to MeV energetic protons } 
\tablenotetext{f} {Two tests for local Diff. Shock Accel (DSA) at shock or compression: velocity gradient, common spectrum for H, He, O, Fe} 
\tablenotetext{g} {One test for non-local DSA: similar spectral slope for H, He} 
\tablenotetext{h} {Four tests for flare source: antisunward anisotropy, hard spectra, high He/H ($>0.004$), high Fe/O ($> 0.124$)   } 
\tablenotetext{i} {Three tests for trapping/confinement: enhancement from suprathermal to EPs, weak anisotropy, lower spectral indices from cooling}
\tablenotetext{j} {Four tests for non-local acceleration, not DSA: no common spectrum for species,  anisotropy, lower He/H, lower Fe/O} 
\end{deluxetable}

 \subsection{Implications for Wave-Particle Acceleration}
 
The work of \cite{Mitchell:2020}  likens the energetic particle population produced in high current density field-aligned current structures with auroral phenomena in planetary magnetospheres. The auroral pressure cooker mechanism is driven by the displacement of the magnetic field and the development of large-currents in these field aligned structures. In the March 2, 2022 event, we have many of the ingredients for the pressure cooker. The CME displaces field lines, and PSP traveled through the flank of the CME where large currents within field structures exist, particularly in regions close to the Sun. 

Despite searching for enhancements in the turbulence and waves, we instead find at best the reductions in inertial range turbulence power where the field and plasma show characertistics of expansion. Local acceleration does not therefore arise from prominent wave-particle acceleration. 

The closed flux tube CME acts to contain energetic  and suprathermal ions. We observe the remnants of acceleration that occured closer to the Sun after these particle distributions have cooled through the expansion of the CME. 

The helicity signatures within the CME flank are quite apparent (see Figure \ref{fig:hmpvi}) as they were in the CME leg observed by \cite{McComas:2023} where an \emph{absence} of energetic particle fluxes were observed. The difference may be associated in part with the interaction. Both the presence of a turbulent sheath and the separated flux tube are indicative of interaction between the CME flank and the surrounding plasma. 

It is possible that EMIC waves were excited closer to the Sun. The elevated levels of PVI in the CME interval (C) are indicative of current sheets or other discontinuities. However, the lack of signatures of signficant changes in the turbulence rules out strong wave-particle interactions. 




 \subsection{Connecting Imaging and \emph{In Situ} Measurements of Flux Rope CME to Particle Acceleration}

The observations of the March 2, 2022 CME provide a way to connect remote and \emph{in situ} measurements. 
We have measured the substructure in situ tied to macro-structure imaged demonstrating the central importance of flux rope in both driving the CME and creating the helicity channels necessary for rapid particle acceleration. 

Remote observations show that the event is associated with a CME, clearly showing a flux-rope morphology with  striated features. The closed configuration of the flux rope is confirmed by strong signatures of bi-directional electron streaming. And the flux-rope morphology is connected to high PVI signatures in the \emph{in situ} measurements, revealing the presence of current-sheets, small-scale structures, and discontinuities. The V-shaped structure observed remotely by WISPR that marks the end of the  CME  and post-loop arcade both indicate high current sheet regions, and the presence of reconnection that reconfigures the magnetic structure. 

It is  likely that at least some of the particle acceleration occurs locally, but not through wave-particle interactions.  The existence of a convective electric field and a large-scale field structure with strong curvature and gradients provides for rapid energization (See \S A.2). However, the mechanism relying only on gradient drift provides an acceleration rate that is too slow.  The PVI and helicity indicate additional likely sources of  energetic particle channeling and confinement.  This notion is consistent with observations of plasma boundaries being associated with energetic particles \cite[]{Pecora:2021}. The broader statistical problem of the association of ion diffusion and acceleration in plasma turbulence is being actively investigated \cite[e.g., ][]{pecora2019statistical, pecora2018ion}. 

Another important component of the observation of energetic particles within the CME  is the high elevated Fe/O and He/H rations. Flaring  contributes to the populations observed in the sheath of the CME, and likely feeds the particle populations into the  CME as well.  There is an association of sympathetic flaring with the eruption of CMEs, and subsequent activity.  Observations  suggest that emerging magnetic flux can trigger both a CME and an associated flare.

In the March 2, 2022 event,  the presence of flaring populations can be justified by the rapid heating of filament material. There is some indication of a loop-top flare arcade in the EUVI 284A images that is consistent with the release of flare particle populations and would explain the observed compositional enhancements in Fe/O and He/H .

\section{Conclusions}
\label{sec:conclude}

We have analyzed Parker Solar Probe  observations on March 2, 2022 when the spacecraft passed through a series of plasma structures at the flank of a small streamer-blowout CME with a flux-rope morphology. The event is a rare case where imaging and \emph{in situ} measurements coincide,  providing complementary information about the plasma in the low corona during the CME ejection and acceleration.  

The first of the intervals observed is a magnetic flux tube that was displaced by the accelerating and expanding CME. Between the CME and the flux tube we observe a turbulent region of solar wind that builds up as a sheath in front of the CME flank. Within the CME, we observe evidence of plasma expansion in the field strength that drops for four hours after passage through the CME interface, and reduced plasma temperature and wind speed. 

The helicity signatures within the CME  are consistent with the flux rope morphology identified in imaging measurements, complete with a V-shaped tail-end of the flux-rope CME.    Both the presence of a turbulent sheath and the separated flux tube are indicative of interaction between the expanding flux rope in the CME  and the surrounding plasma. 

Despite the absence of a shock or strong compression within the CME, we observe rich SEP composition including enhancements in He/H and Fe/O. Anisotropies indicate nearly isotropic distributions for $\sim$100 keV protons. 
The spectra  of H and He indicate a common spectrum above 1 MeV/nuc, but distinct slopes below 1 MeV/nuc with protons having the softest spectra. 

\begin{figure}[ht]
    \centering
    \includegraphics[width=\textwidth]{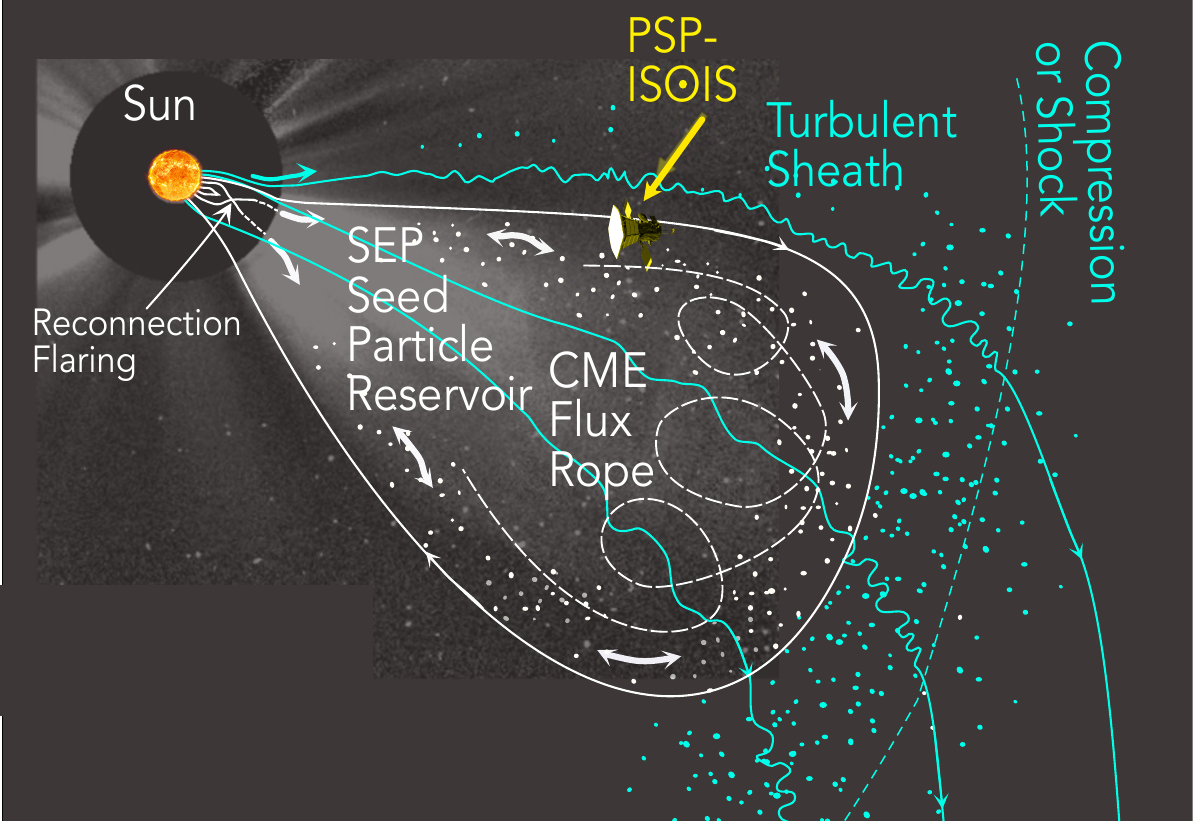}
  \caption{The observations of particle populations within the CME show that it acts as a reservoir, building up seed populations rich in energetic particles from flares and accelerated plasma. The CME  overtook PSP on March 2, 2022 revealing the connection between the closed flux tubes, the aniostropies and the energetic particle seed populations stored within the SBO flux-tube CME. }
  \label{fig:f14}
  \end{figure}

A search for turbulent field changes in the CME shows no evidence of strong enhancements in the power-spectral densities. In fact, the strongest changes in the power-spectra are observed as reductions in regions of plasma expansion in both the CME  and the flux rope. 

Together these observations show containment of energetic particles below 1 MeV/nuc within the flank of  the CME. The  appearance of He/H and Fe/O enhancements in the energetic particle populations suggests that flares feed particles into the CME, which acts as a reservoir and builds up high fluxes of energetic particles.  As the CME propagates out through the inner heliosphere, it drives compression and potentially a shock across the CME front where energetic particles are diffusively accelerated. 

The observations within the CME indicate that the enhancments in particles' differential fluxes are connected from the suprathermal  through the energetic particle energies ($\sim$ keV to $>10$ MeV ), showing that energetic particles behave as a connected population and are mostly contained within the CME. However, the absence of significant enhancements in the turbulent power-spectra rules out significant wave-particle acceleration as the dominant accleration mechanism, at least in the flank of the CME where PSP moved through the structure. The combination of generally high levels of turbulence, the large field strengths close to the Sun,  and large helicity from the field curvature likely create helicity channels that locally confine energetic particles. This notion is consistent with observations of plasma boundaries being associated with energetic particles \cite[]{Pecora:2021}.  The consistency between imaging observations of the flux rope morphology and the \emph{in situ} observations showing   small-scale flux-rope structures reveal that the flux ropes internal to the CME play a central role in not only the acceleration of the CME, but also in confining, and channeling particles. 

CMEs are known to interact with the surrounding solar wind, driving magnetic reconnection \cite[]{Winslow:2016}, and progressively opening through interchange reconnection \cite[]{Schwadron:2010}. Therefore, the energetic particle reservoir contained by the CME is progressively emptied into the compressive or shocked  sheath in front of the CME (Figure \ref{fig:f14}). The CME reservoir not only retains energetic particles, but releases these distributions out into the inner heliosphere where stronger compression and shocks form naturally. Therefore, CMEs act to retain and then deliver energetic particle seed populations to the compressions and shocks where they are efficiently accelerated to higher energies. 

Thus, we reveal energetic particle populations together with remote and \emph{in situ} plasma measurements of the flank of a coronal mass ejection, the associated sheath, and nearby plasma structures. The observations reveal that the CME acts as an energetic particle resorvoir. The flux-rope morphology of the CME helps to contain energetic particles, particularly along helicity channels and other  plasma boundaries. The CME builds-up  energetic particle populations, allowing them to be fed into subsequent higher energy particle acceleration throughout the inner heliosphere where a compression or shock forms on the CME front.  The synthesis reported here is enabled by the unique  PSP/\ISOIS~  energetic particle observations together with the \emph{in situ} and remote PSP plasma observations close to the Sun. 

\acknowledgments

We are deeply indebted to everyone who helped make the
Parker Solar Probe (PSP) mission possible. We thank all of the
outstanding scientists, engineers, technicians, and administrative
support people across all of the \ISOIS~, FIELDS,  SWEAP, and WISPR
institutions who produced and supported the  instrument suites, support its operations and the scientific
analysis of its data. This work was supported as a part of the PSP
mission under contract NNN06AA01C. The \ISOIS~ data and
visualization tools are available to the community at: https://
spacephysics.princeton.edu/missions-instruments/isois; data are
also available via the NASA Space Physics Data Facility (https://
spdf.gsfc.nasa.gov/). Parker Solar Probe was designed, built, and
is now operatedby the Johns Hopkins Applied Physics Laboratory
as part of NASA’s Living with a Star (LWS) program (contract
NNN06AA01C). Support from the LWS management and
technical team has played a critical role in the success of the
Parker Solar Probe mission.

\appendix

\section{Particle Energization from Flux Tube Disruption}

Observations from Parker Solar Probe on  March 2, 2022 indicate a brief period with a twisted flux tube that becomes distorted through interaction near the flank of a CME. Appendix A.1 considers the flux tube properties  and its interaction with the CME. 	Appendix A.2 discuss particle acceleration across plasma boundaries with the CME and the flux tube. 

\subsection{Field Stress and Pressure Balance across the March 1, 2022 Small Flux Tube} 

Consider first the large-scale properties of the flux tube observed between DOY 60 (March 1, 2022)  between times 21:05:27.270 and 23:44:08.319. We observe a large decrease in density and temperature, while the magnetic field strength shows a small increase. The field components indicate a slow rotation (for example, the normal component increases by 50 nT and then returns to its previous value). A first question is how this structure remains in pressure balance with its surroundings given then large decrease in internal pressure. 

The pressure gradient across the flux tube and the the $-\mathbf{j} \times \mathbf{B}$ forces (field pressure and field-line tension) contribute to the force densities exerted by the flux rope on the surrounding solar wind plasma:
\begin{eqnarray}
- \nabla p - \nabla \frac{B^2}{8\pi} + \frac{\mathbf{B} \cdot \nabla \mathbf{B}}{4\pi} \approx \mathrm{Force~ density}
\end{eqnarray}
In this case, the field line tension appears to dominate over the change in field pressure across the small flux tube structure. (We return to this point when considering cross-field particle drift.) Given a propagation speed of $u_\mathrm{sft} \sim 250$ km s$^{-1}$ and the time duration over which the flux tube is observed $\tau_\mathrm{sft} \approx 2.64$ hrs (the subscript `sft' indicates `small flux tube'). This suggests that the structure has a width of at least 0.015 au and a minimum gradient scale of $\lambda_\mathrm{sft} \sim 0.008$ au.  Note the large pressure change from outside to within the structure of $\Delta p_\mathrm{sft} \sim 3400$ eV cm$^{-3}$. The associated pressure gradient $\Delta p_\mathrm{sft} /\lambda_\mathrm{sft}$ exerts a force density directed inward toward the interior of the flux rope, whereas the field-line tension is directed outward from the flux rope.  

The observed tensional force density  $\sim B_N^2/(4\pi \lambda_\mathrm{sft})$ is approximately $3.7 \times$ the  pressure gradient. This indicates that the flux tube is not in equilibrium. The $\mathbf{j} \times \mathbf{B}$ forces of the flux tube drive its expansion as it is overtaken by structures associated with the coronal mass ejection behind it. This outward expansion of the flux rope appears to explain the lower plasma pressure within the structure.

\subsection{Energization through Cross-Field Transport near Plasma Boundaries and Flux Tubes}
\label{sec:xfield}

The presence of energetic particles within the expanding CME  appears at first as a contradiction. Expanding structures should cool as they expand. In this sense, the pressure drop in the interior of the CME is as expected, but the presence of accelerated particles suggests some form of interaction. 




An important consideration in this interaction is the presence of a convective electric field in the frame of the sheath. Note that the trailing edge of the flux rope shows compression in the magnetic field. Presumably this is associated with sheath material that slows down as it overtakes the flux rope. Similarly, we observe an enhancement in the field strength near the beginning of the CME interval, again indicating interaction and compression of the CME  near its interface with the sheath. The presence of energized particles are observed throughout the region of compression. Particles drift across the magnetic field in this region of compression and acquire energy directly as they move in the direction of the convective electric field.

Any acceleration making use of the convective electric field requires the presence of a change in speed ($\Delta u$). Stated differently, if the plasma all moved at a uniform speed, then we can transform into the frame of reference of the plasma where the convective electric field vanishes and cannot cause particle acceleration. In this sense, the signatures of compression are essential in showing that a differential change in plasma speed likely existed upstream, closer to the Sun.  Further, in examining Figure 5, we observe multiple speed changes: Interval A shows leads with a slower speed and trails with a larger speed, Interval B shows higher speeds in general with significant speed fluctuations throughout the interval, and Interval C generally shows speed reductions. Notably, the large changes in speed are at or near the plasma boundaries separating these intervals. More generally, even in the presence of  small changes in  plasma velocity, the convective electric field is a 3D time varying function. Large variations in this field occur where the magnetic field varies strongly relative to the plasma flow. For example, a magnetic field parallel to the bulk flow produces a vanishing convective electric field, whereas a strong magnetic field perpendicular to the bulk flow produces a large convective electric field. Therefore, boundaries between magnetic flux tubes will typically produce large changes in the convective electric field due to strong variations in the magnetic field.  

Referring back to our discussion of PVI in  \S 2.2, we note that large PVI values were associated  with the highly turbulent environment in Interval B, and with boundaries of flux tubes. Similarly,  in Interval C, we found evidence of flux tubes  on multiple scales, and the  large-PVI values are found at the boundaries of large (in magnitude) helicity events. The large PVI values are not necessarily near large speed gradients, but they do appear either in more turbulent periods or near plasma boundaries where strong variations in the convective electric field is expected.

A similar acceleration mechanism invoking the convective electric field was invoked by \cite{Jokipii:1982} for the acceleration of anomalous cosmic rays at the termination shock. In this case, the acceleration mechanism involves particle drift in the highly transverse fields of the outer heliosphere, and the convective electric field is associated with the relative motion of the solar wind as it traverses the shock. The difference  with the plasma structures considered here is that the acceleration region is small and the fields are strong, whereas the drift near the termination shock occurs over enormous spatial regions where the magnetic fields are far weaker.  In the case of the termination shock, the acceleration occurs just upstream of the shock, and the solar wind moving relative to the shock drives the convective electric field.  In the case of the plasma structures observed here, the compression exists on the trailing edge of the flux rope, within the sheath, and in the CME near its interface with the sheath. It is the flux rope and the CME moving with respect to the sheath that drive the convective electric field.

The speed changes are  $\Delta u \sim 150$ km s$^{-1}$ toward the sheath, opposite the radial direction in the flux rope, or in the radial direction in the CME.   The convective electric field $\mathbf{E}_c  = - \Delta \mathbf{u} \times \mathbf{B} / c$  exists within the compressed field.  Near the boundaries of these structures, the field is dominated by its radial component. However, interior to these structures, the tangential component in the field becomes large. If we equate the  convective electric field with a potential $\mathbf{E}_c = - \nabla \phi_E$, the electric potential inside the structures increases in the Normal direction in RTN coordinates. This potential is quite large: given a gradient scale of 0.008 au, a tangential field of -50 nT and relative flow speed of 150 km s$^{-1}$, a charged particle could gain $\sim 9$ MeV/Q by drifting through the convective electric field toward the interior. This estimate takes into account the field strength observed locally. However, the magnetic field is significantly stronger closer to the Sun (at smaller radial distances the field increases as at least $r^{-2}$). Therefore, the available energy for acceleration is significantly larger than 10 MeV/Q. 

Curvature and gradient drift is expressed as follows,
\begin{eqnarray}
\mathbf{v}_d & = &  \frac{c v p}{3 q} \nabla \times \frac{\mathbf{B}}{B^2}  
\label{eq:drift}
\end{eqnarray}
The curvature drift interior to the flux rope is approximately directed in the -T direction (opposite of the tangential direction), and the gradient drift is approximately in the -N direction. The drifts are not as ordered in the CME. We take the drift in the flux tube  as a representative example of the substantial energy gain that can be realized through cross-field drift, but also the inherent problems in the mechanism. The curvature drift moves positively charged ions, on average, in a direction perpendicular to the convective electric field, resulting in small, if any, energy gain. The gradient drift moves ions, on average, toward decreasing potential, and thereby increases the energy of these particles, as observed. 

We encounter several problems if the effect is the result of drift. The first is that the characteristic drift speed is quite small. For a $\sim$100 keV/nuc He ion (doubly charged alpha particle), in a 90 nT magnetic field strength, and given a gradient scale of 0.008 au, the drift speed is $\sim$  2.5 km s$^{-1}$. For an ion to traverse the gradient through drift would require more than 5 days, which far exceeds the $\sim 1.5$ day propagation of the structure to 0.22 au. The second issue is that this formulation of drift assumes a nearly isotropic distribution function, whereas the observed distribution is highly anisotropic. It is possible that channels of cross-field drift and cross-field diffusion occur more rapidly over portions of the turbulent magnetic field where the curvature in the field is large, or the magnetic field strength is low. In other words, the drift may happen on much smaller kinetic scale lengths. 
 
Within the small flux tube, the large anisotropy indicates that particles are more likely from flares or particles accelerated close to the Sun that propagate out along the magnetic field. While local energization of the ions is not ruled out, the large anisotropies observed point to a flare source and a  non-local acceleration process.  

In contrast to the small flux tube,  the flank of the CME shows particle containment. Throughout this region, we see flux tubes and gradients in the magnetic field strength. Particle motion within these flux tubes causes repetitive energy gains until the gyroradius exceeds the size of the flux tube, allowing the particle to escape the acceleration region. Flux tubes within the CME are observed for typically 20 min, indicating diameters of $\sim 0.0024$ au (or 0.51 R$_s$). At these field strengths, the particle gyroradius of a proton does not exceed the diameter of a typical flux tube until the particle energy exceeds 5.6 GeV.  The acceleration process is likely limited by the rate of acceleration, and the propagation time of the flux tubes to 0.22 au.

\bibliographystyle{apj} 

\end{document}